\documentclass[ sn-apa]{sn-jnl}% Default
\usepackage{graphicx}%
\usepackage{multirow}%
\usepackage{amsmath,amssymb,amsfonts}%
\usepackage{amsthm}%
\usepackage{mathrsfs}%
\usepackage[title]{appendix}%
\usepackage{xcolor}%
\usepackage{textcomp}%
\usepackage{manyfoot}%
\usepackage{longtable}
\usepackage{booktabs}%
\usepackage{algorithm}%
\usepackage{algorithmicx}%
\usepackage{algpseudocode}%
\usepackage{array}
\usepackage{listings}%
\usepackage{rotating}
\usepackage{makecell}
\usepackage{bm}%
\usepackage{lineno}
\usepackage{hyperref}
\usepackage {natbib}
\usepackage{threeparttable}
\usepackage{geometry}   %设置页边距的宏包
\usepackage{titlesec}   %设置页眉页脚的宏包
\theoremstyle{thmstyleone}%
\theoremstyle{thmstyletwo}%

\theoremstyle{thmstylethree}%

\begin{document}

\title[Article Title]{Toward Pedestrian Head Tracking: A Benchmark Dataset and a \textcolor{black}{Multi}-source Data Fusion Network}
% \title[Article Title]{Toward a Large-scale Cross-scene Pedestrian Head Tracking Dataset and Multi-Source Information Fusion Network in Dense Crowds}


\author[1,3]{\fnm{Kailai} \sur{Sun}}\email{skl24@mit.edu}

\author[2]{\fnm{Xinwei} \sur{Wang}}\email{wxw21@tsinghua.org.cn}

\author*[4]{\fnm{Shaobo} \sur{Liu}}\email{shaobo@whut.edu.cn}

\author[2]{\fnm{Qianchuan} \sur{Zhao}}\email{zhaoqc@tsinghua.edu.cn}

\author[2]{\fnm{Gao} \sur{Huang}}\email{gaohuang@tsinghua.edu.cn}

\author[4]{\fnm{Chang} \sur{Liu}}\email{liuchang.20@whut.edu.cn}

% 0000-0003-1648-3409(Kailai Sun);0000-0001-9104-6503(Xinwei Wang);0000-
% 0002-7952-5621(Qianchuan Zhao).
\affil[1]{\orgdiv{Singapore-MIT Alliance for Research and Technology Centre (SMART)}, \orgname{Massachusetts Institute of Technology}, \orgaddress{\postcode{138602}, \country{Singapore}}}

\affil[2]{\orgdiv{Department of Automation}, \orgname{Tsinghua University}, \orgaddress{ \city{Beijing}, \postcode{100084}, \country{China}}}

\affil[3]{\orgdiv{Urban Mobility Lab}, 
          \orgname{Massachusetts Institute of Technology}, 
          \orgaddress{\city{Cambridge}, \state{MA 02139}, \country{USA}}}

\affil[4]{\orgdiv{Intelligent Transportation Systems Research Center}, \orgname{Wuhan University of Technology}, \orgaddress{\city{Wuhan Hubei}, \postcode{430063}, \country{China}}}

\abstract{Pedestrian detection and tracking in crowded video sequences have many applications, including autonomous driving, robot navigation and pedestrian flow analysis. However, detecting and tracking pedestrians in high-density crowds face many challenges, including intra-class occlusions, complex motions, and diverse poses. Although artificial intelligence (AI) models have achieved great progress in head detection, head tracking datasets and methods are extremely lacking. Existing head datasets have limited coverage of complex pedestrian flows and scenes (e.g., pedestrian interactions, occlusions, and object interference). It is of great importance to develop new head tracking datasets and methods. To address these challenges, we present a Chinese Large-scale Cross-scene Pedestrian Head Tracking dataset (Cchead) and a Multi-source Data Fusion Network (MDFN). The dataset has features that are of considerable interest, including 10 diverse scenes of 50,528 frames with about 2,366,249 heads and 2,358 tracks. Our dataset contains diverse pedestrian moving speeds, directions, and complex crowd pedestrian flows with collision avoidance behaviors. Existing state-of-the-art (SOTA) algorithms are tested and compared on the Cchead dataset. MDFN is the first end-to-end convolutional neural network (CNN)-based head detection and tracking network that jointly trains Red, Green, Blue (RGB) frames, pixel-level motion information (optical flow and frame difference maps), depth maps, and density maps in videos. \textcolor{black}{Ablation experiments confirm the significance of multi-source data fusion.} Compared with SOTA pedestrian detection and tracking methods, MDFN achieves superior performance across three datasets: Cchead, Restaurant and \textcolor{black}{Crowd of Heads Dataset (CroHD)}. To promote further development, we share our source code and trained models for global researchers: https://github.com/kailaisun/Cchead. We hope our datasets to become essential resources towards developing pedestrian tracking in dense crowds.}

\keywords{Deep learning, Pedestrian detection and tracking, Head tracking, Information fusion, Crowded scenes}



\maketitle

\clearpage

% \begin{table}[ht]
% \centering
% \begin{tabular}{|p{2cm}|p{5cm}|}
% \hline
% \textbf{Acronym} & \textbf{Definition} \\ \hline
% CNN & Convolutional Neural Network \\ \hline
% CPU & Central Processing Unit \\ \hline
% DNN & Deep Neural Networks \\ \hline
% GPU & Graphics Processing Unit \\ \hline
% MLP & Multilayer Perceptron \\ \hline
% RAM & Random Access Memory \\ \hline
% SVM & Support Vector Machine \\ \hline
% ViT & Vision Transformer \\ \hline
% \end{tabular}
% \caption{List of Acronyms}
% \label{tab:acronyms}
% \end{table}

% \linenumbers 

\section{Introduction}\label{sec1}
Pedestrian detection and tracking in crowded video sequences are central problems in the understanding of visual scenes, with widespread applications including autonomous driving \citep{QIAN2022108796,MO2023105705}, video surveillance \citep{wang2018resource,SUN2022109354}, pedestrian flow dynamics \citep{AGHAMOHAMMADI202099}, and robot navigation \citep{WOS:000939288300012}. The complexity of this task increases with the density of pedestrians to be tracked. The recent focus on the Multiple Object Tracking Challenge (MOTChallenge) benchmark has shifted toward tracking pedestrians in high-density crowds \citep{10.1007/978-3-030-58558-7_26,8953401,9880207,10030387,Dendorfer2020MOT20AB}. With the rapid development of deep learning and large language models (LLMs), convolutional neural networks (CNNs) and Transformer methods have become mainstream \citep{TSAI2023105770,LI2023106527}. 

However, detecting and tracking pedestrians in high-density crowds poses significant challenges, including intra-class occlusions \citep{HU2024112130,FUENTESJIMENEZ2021104484}, complex motions, and diverse poses. Thus, with pedestrian density increases, the visibility of pedestrians decreases, resulting in decreased pedestrian detection and tracking accuracy. Although existing pedestrian detection and tracking datasets and methods \citep{9711285,Hasan_2021_CVPR, Stadler_2021_CVPR,10400895} have achieved remarkable progress, the bounding box-level tracking performance is saturating.
% In recent years, most on-line tracking algorithms follow the paradigm of detection and tracking combination, and several studies have found that the performance of the object detector is crucial for the performance of the tracker.

To tackle these problems and effectively detect and track pedestrians in dense and crowded scenes, researchers have turned to the clear and visible part of the human: head. Although existing datasets and methods are available for head detection \citep{DBLP:journals/corr/abs-1809-03336,liu2021head,liu2021samnet,DBLP:journals/corr/abs-1809-08766,shen2019indoor,ISI:000521828604112,10400895}, there is an extreme scarcity of datasets and methods specifically dedicated to head tracking \citep{9577483}. Besides, existing head detection and tracking datasets suffer from limited data volume and diversity, significantly limiting researchers' ability to develop data-hungry artificial intelligence (AI) models that require large-scale tracking datasets. Existing head detection and tracking datasets have limited coverage of complex pedestrian flows and scenes (e.g., pedestrian interactions, occlusions, and object interference).

One of the main reasons for the scarcity of large-scale head tracking datasets is the high cost of their collection and annotation \citep{shao2018crowdhuman}. Unlike classification or detection datasets, the collection and annotation processes of the head tracking dataset involved are non-trivial and significantly more complex, as both data curation (sourcing) and annotation demand substantially more manual human labour. For head tracking, annotation is challenging because of the smaller size of the head bounding boxes, compared with people tracking. For head tracking in a crowd, annotating and tacking each head accurately is particularly challenging because of the frequent occlusions. The small size of head boxes further increases the likelihood of errors, as heads are easily occluded by other individuals. Besides, the small and indistinct head regions make it hard for annotators to consistently follow head movements, leading to potential tracking inaccuracies, especially in densely populated scenes.

To address the above challenges, we introduce a new Chinese large-scale cross-scene head tracking dataset in dense crowds, named Cchead. Our dataset consists of 10 scenes, including nature pedestrian flow scenes and controlled pedestrian flow experiment scenes at several different types of locations, such as traffic, buildings, and schools, with 2,366,249 manually annotated head bounding boxes for tracking purposes. Cchead consists of 50,528 frames and 2,358 tracks annotated in diverse scenes. Cchead contains diverse pedestrian moving speeds, directions, and complex crowd pedestrian flows with avoidance situations. The number of pedestrians in Cchead ranges from 11 to 89 people per frame. Since existing mainstream head detection and tracking methods encompass two perspectives (Slope and Overhead views), Cchead provides videos from both perspectives to facilitate a wider range of application scenarios.

In addition, although existing methods for detecting and tracking pedestrian heads have made advancements, the tasks are still challenging. In crowded scenes, background objects frequently share similar features with human heads—such as colour, size, and texture, resulting in the difficult detection and tracking of small-scale, variably posed, and low-light heads. Additionally, the small size of these heads makes it challenging to detect and track them with high confidence scores.  Furthermore, the movement of heads introduces diversity in scale, pose, and texture.

The input information source is critical to achieving accurate head detection and tracking. Motion information (e.g., optical flow and frame difference) can enhance head features and suppress background features. Depth maps can provide effective distance information, and density maps can highlight the spatial features of heads. Existing research mainly focuses on head detection in static images, depth maps, or optical flow. They did not consider how to integrate multiple sources of information to address the aforementioned challenges.

Inspired by these observations, this study proposes a Multi-Source Data Fusion Network (MDFN) for pixel-level head tracking. This is the first end-to-end video-based network that jointly trains raw Red, Green, Blue (RGB) frames, pixel-level motion data (optical flow and frame difference maps), depth maps, and density maps. To obtain the five-source information, we adopt a transfer learning strategy without external sensors. MDFN leverages the five-source data to effectively guide the head feature extraction, resulting in robust head detection and tracking. We have developed a feature fusion mechanism that extracts a comprehensive latent representation, which combines the strengths of each input source for head tracking. \textcolor{black}{This study provides a large-scale cross-scene dataset and a novel multi-source data fusion insight for head detection and tracking. This is rarely seen in existing literature. The dataset is collected from real-world scenarios, making it highly applicable in practical applications while also providing researchers with broader opportunities for further exploration. The proposed multi-source data fusion approach is detector-free, so it can be compatible with various feature extraction methods and has the potential to be adopted and applied in a wider range of SOTA detection and tracking networks.}

% \begin{table*}[ht!]
% \caption{Cchead dataset statistics}
% \vspace{-0.5em}
% \begin{center}
% \begin{tabular}{@{}cccccccccc@{}}
% % \begin{tabular}{lllll} %l(left)居左显示 r(right)居右显示 c居中显示
% \toprule
% Scenario&Views&Boxes&Fps&Time&Frames&Density&Tracks & Mean Area & Mean Speed\\
% \midrule 
% Classroom&Slope&61,884&30&45s&1,452&42.62&254 & 1.57e-3 & 0.0566\\
% Roof(+)&Slope& 225,816&50&50s&2,531&89.22&114 & 1.46e-4 & 0.1045\\
% Roof(Y)& Slope&191,101&50&45s&2,275&84.00&90 & 1.37e-4 & 0.1082\\
% Office& Slope&46,965&~&~&4,178&11.24&16 & 1.41e-3 & 0.1489\\
% Roof(T)& Slope&170,869&50&40s&2,002&85.35&90 & 1.60e-4 & 0.1108\\
% Street& Overhead&166,000&25&203s&5,083&32.66&468 & 3.44e-4 & 0.3001\\
% School Road 1&Slope&512,942& 50&220s &11,002& 46.62& 400 & 2.17e-4 & 0.1827 \\
% School Road 2&Slope&360,534&50&220s&11,001&32.77& 260 & 2.15e-4 & 0.1075 \\
% School Parking Lot 1& Overhead& 431,548&50& 140s &7,001&61.64& 437 & 1.15e-4 & 0.2106 \\
% School Parking Lot 2&Overhead&198,590& 50FPS&80s &4,001& 49.63& 229 & 1.09e-4 & 0.2280\\
% Total& ~&2,366,249&~&~&50,528&46.83&2,358 &  &  \\
% \bottomrule
% \end{tabular}
% \vspace{-0.5em}
% \label{dataset}
% \end{center}
% \end{table*}

\begin{table*}[ht!]
\caption{Cchead dataset statistics}
\vspace{-0.5em}
\begin{center}
\begin{tabular}{@{}lllllllll@{}}
% \begin{tabular}{@{}p{3.5cm}p{1.2cm}p{1.5cm}p{0.3cm}p{0.5cm}p{1cm}p{1cm}p{0.8cm}p{1cm}@{}}
% \begin{tabular}{lllll} %l(left)居左显示 r(right)居右显示 c居中显示
\toprule
Scenario&Views&Boxes&Fps&Time&Frames&Density&Tracks &  Speed\\
\midrule 
Classroom&Slope&61,884&30&45s&1,452&42.62&254 &  0.0566\\
Roof(+)&Slope& 225,816&50&50s&2,531&89.22&114 &  0.1045\\
Roof(Y)& Slope&191,101&50&45s&2,275&84.00&90 & 0.1082\\
Office& Slope&46,965&~&~&4,178&11.24&16 &  0.1489\\
Roof(T)& Slope&170,869&50&40s&2,002&85.35&90 &  0.1108\\
Street& Overhead&166,000&25&203s&5,083&32.66&468 &  0.3001\\
School Road 1&Slope&512,942& 50&220s &11,002& 46.62& 400 &  0.1827 \\
School Road 2&Slope&360,534&50&220s&11,001&32.77& 260 &  0.1075 \\
School Parking Lot 1& Overhead& 431,548&50& 140s &7,001&61.64& 437 & 0.2106 \\
School Parking Lot 2&Overhead&198,590& 50&80s &4,001& 49.63& 229 & 0.2280\\
% \midrule 
Total& ~&2,366,249&~&~&50,528&46.83&2,358 &    \\
\bottomrule
\end{tabular}
\vspace{-0.5em}
\label{dataset}
\end{center}
\end{table*}

The key contributions of this study include:

\begin{itemize}
\item We collect and annotate a Chinese large-scale cross-scene dataset (Cchead) for crowded head tracking.
\item We conduct experimental comparison and analysis using existing state-of-the-art (SOTA) multi-object tracking algorithms on our Cchead dataset.
\item We extend the Cchead dataset with multi-source data by a transfer learning strategy.
\item We present a novel insight and deep network (MDFN) for head detection and tracking by leveraging pseudo multi-source data fusion. \textcolor{black}{Ablation experiments confirm the significance of multi-source data fusion.}
\item Experimental results demonstrate MDFN as a strong baseline by comparing it to existing SOTA multi-object tracking algorithms \textcolor{black}{across three datasets: Cchead, Restaurant and CroHD}. To promote further development, we share our source code and trained models for global researchers: https://github.com/kailaisun/Cchead. 
\end{itemize}

% The multi-source information fusion enables the network to leverage complementary cues from different sources, improving the robustness and adaptability of head detection in various scenarios. The use of pixel-level motion information, depth maps, and density maps enhances the network's ability to handle challenging situations such as occlusions, diverse poses, and low-light conditions. The end-to-end training approach ensures that the network learns optimal feature representations for head detection directly from the input data, without relying on pre-processing or post-processing steps.

% In conclusion, this chapter presents a novel approach for video-based head detection using a Multi-Source Information Fusion Network (MDFN). The proposed network integrates RGB frames, pixel-level motion information, depth maps, and density maps to improve the accuracy and robustness of head detection in dense and crowded environments. The experimental results demonstrate the effectiveness of the MDFN network in various scenarios, showcasing its potential for advancing research in pedestrian tracking and understanding dense human crowds.

\section{Related work}
% \subsection{Multi-Object Tracking Datasets} 
\subsection{Head detection datasets}

Head detection task has many challenges, including small scales, diverse poses, similar background objects, and occlusions. Early head detection datasets
\citep{ISI:000380414100323,6909512,6133287} provide ground truth head bounding boxes in Television (TV) movies.

Recent studies have focused on crowded scenes: the South China University of Technology (SCUT)-HEAD dataset \citep{8545068} includes
4,405 images with 111,251 heads annotated; Brainwash crowd dataset \citep{DBLP:journals/corr/StewartA15} includes 11,917 images with 91,146 heads annotated; Restaurant dataset \citep{9150687} includes 1,610 images with 16,060 heads annotated; CrowdHuman dataset \citep{shao2018crowdhuman} includes 24,370 images with 470K heads annotated; 
Microsoft Common Objects in Context (COCO) HumanParts includes 66,808 images with 232,392 heads annotated; \textcolor{black}{ GigaPixel-level humAN-centric viDeo dAtaset (PANDA)  includes 21 real-world outdoor scenes, 127k human trajectories, and 4k head annotated \citep{9156646}. Railway Platforms and Event Entrances-Heads (RPEE-Heads) dataset includes 109913 head annotations across 1886 images \citep{abubaker2024rpee}.}
Crowd of Heads Dataset (CroHD) dataset \citep{9577483} attempts to annotate head tracking chains from an existing pedestrians tracking dataset \citep{MOT19_CVPR}, including 11,463 frames with 2,276,838 heads annotated.

\subsection{Head detection methods}
% Most related studies treat head detection as a subtask of object detection. Early methods extracted hand-crafted features (Haar features \citep{INSPEC:7176899}, 
% HOG features \citep{RN475}, etc.) and utilized classifiers to detect heads. With the advent of Deep Learning, Convolutional Neural Networks (CNN) and Transformer methods have become
% mainstream \citep{DBLP:journals/corr/StewartA15,chouai2021new,khan2020robust,babu2017switching,WOS:000870283000011,ccc1o9}. 

Head detection has traditionally been considered a specialized subtask in object detection (OD). Early approaches primarily relied on hand-crafted features such as Haar-like features \citep{INSPEC:7176899} and Histogram of Oriented Gradients (HOG) features \citep{RN475}. These features were fed into classifiers like Support Vector Machines (SVM) or AdaBoost to detect heads in images. With the development of deep learning, methodologies shifted towards leveraging CNNs and Transformer-based architectures, which have become the mainstream in object detection tasks \citep{DBLP:journals/corr/StewartA15,chouai2021new,khan2020robust,babu2017switching,ccc1o9}. These deep learning models automatically learn hierarchical feature representations, significantly improving object detection accuracy beyond traditional methods.

However, pedestrian head detection presents unique challenges distinguishing it from generic object detection. Pedestrian heads often occupy a small portion of the image, especially in crowded scenes, and can vary greatly in appearance due to occlusions, diverse poses, and lighting conditions. Unlike generic objects, heads are more susceptible to being missed or misclassified. Thus, \textcolor{black}{the multi-scale features aggregation mechanism \citep{Wu_2023_CVPR, vo2022pedestrian, xiang2017joint}}, specific anchor size selection mechanism~\citep{DBLP:journals/corr/abs-1809-08766,Khan2021},  image pyramids mechanism \citep{8545068}, attention selection mechanism \citep{ccc1o9,li2019dsfd,shen2019indoor}, motion enhancement mechanism \citep{SUN2022109354}, \textcolor{black}{ count-aware mechanism \citep{10745131}} and the head-body matching mechanism \citep{10400895,RN609} can effectively improve detection performance. 

\begin{figure*}[htbp]
% \begin{center}
 \begin{minipage}{0.32\linewidth}
  \centerline{\includegraphics[width=\textwidth]{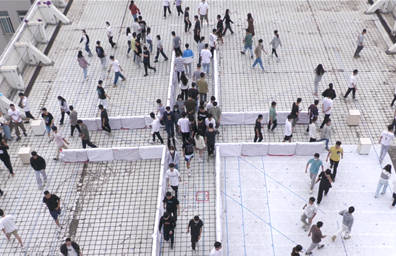}}
 \end{minipage}
 \begin{minipage}{0.32\linewidth}
  \centerline{\includegraphics[width=\textwidth]{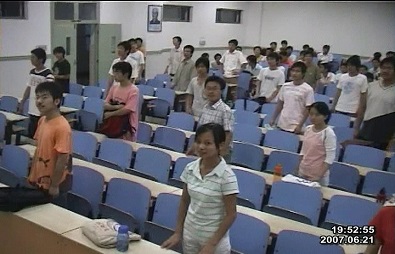}}
 \end{minipage}
 \begin{minipage}{0.33\linewidth}
  \centerline{\includegraphics[width=\textwidth]{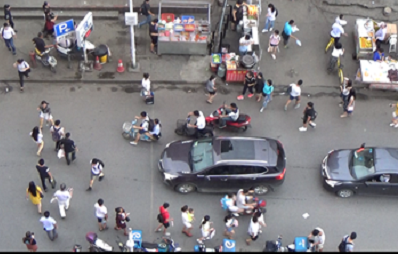}}
 \end{minipage} \\
 \begin{minipage}{0.32\linewidth}
  \centerline{\includegraphics[width=\textwidth]{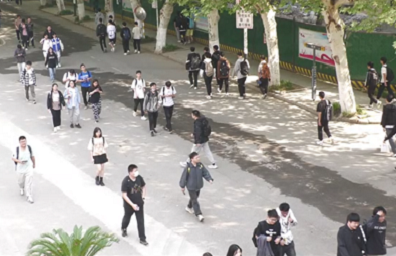}}
 \end{minipage}
 \begin{minipage}{0.32\linewidth}
  \centerline{\includegraphics[width=\textwidth]{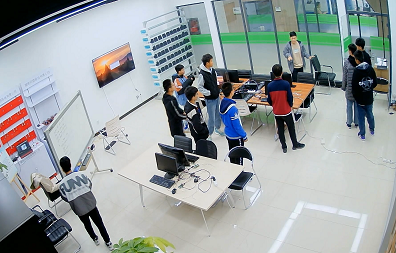}}
 \end{minipage}
 \begin{minipage}{0.33\linewidth}
  \centerline{\includegraphics[width=\textwidth]{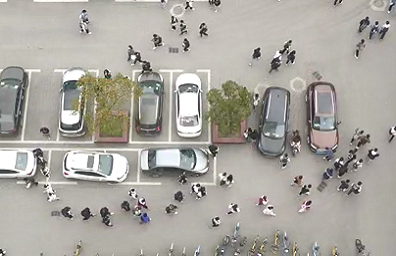}}
 \end{minipage}
\caption{Examples in our Cchead dataset. The scenes include the roof, classroom, street, school road, school parking lot, office, and so on.}
\vspace{-0.5em}
\label{datasetexample}
\end{figure*}

% \begin{figure*}[htbp]
% % \begin{center}
%  \begin{minipage}{0.33\linewidth}
%   \centerline{\includegraphics[width=\textwidth]{c1.jpg}}
%   % \centerline{Image}
%  \end{minipage}
%  \begin{minipage}{0.33\linewidth}
%   \centerline{\includegraphics[width=\textwidth]{c3.jpg}}
%  \end{minipage}
%  \begin{minipage}{0.33\linewidth}
%   \centerline{\includegraphics[width=\textwidth]{c4.jpg}}
%  \end{minipage}

%  \begin{minipage}{0.33\linewidth}
%   \centerline{\includegraphics[width=\textwidth]{c7.png}}
%  \end{minipage}
%  \begin{minipage}{0.33\linewidth}
%   \centerline{\includegraphics[width=\textwidth]{c5-2.jpg}}
%  \end{minipage}
%  \begin{minipage}{0.33\linewidth}
%   \centerline{\includegraphics[width=\textwidth]{c6.jpg}}
%  \end{minipage}
% \caption{Examples in our Cchead dataset. The scenes include the roof, classroom, street, school road, school parking lot, office, and so on.}
% \vspace{-0.5em}
% \label{datasetexample}
% \end{figure*}

\subsection{Multi-Object Tracking (MOT) methods} 

MOT focuses on tracking objects in an image sequence or a video \citep{LUO2021103448,8953401}. MOT can be divided into two frameworks \citep{WANG2023107084}: separate detecting and embedding (SDE) and joint detecting and embedding (JDE). SDE splits the MOT tasks into two independent subtasks: object detection and re-identification (ReID). Recent SDE MOT methods achieve high performance: deep simple online and realtime tracking (DeepSORT) \citep{wojke2017simple}, Bytetrack \citep{BS6-WOS:000904116000001},  observation-centric (OC)-SORT\citep{li2020observation}, \textcolor{black}{deep OC-SORT \citep{10222576}}, BoT-SORT \citep{aharon2022bot}. The advantage of SDE is that each part can be optimized independently, leading to flexible training and high accuracy. On the other hand, JDE combines object detection and ReID feature extraction into a single network by multi-task learning, which effectively reduces computational complexity and improves inference speeds, including CenterTrack \citep{zhou2020tracking}, {\color{red}DiffMOT \citep{Lv_2024_CVPR}}, FairMOT \citep{BS6-WOS:000692902100001} and JDE \citep{BS6-INSPEC:20258751}. The ReID network can extract aggregated latent high-dimensional features, but the training process will be more difficult.

\section{Cchead Dataset}

% \subsection{Dataset Statistics}
This section will describe how our Cchead dataset is collected and annotated.

We captured videos in 10 scenes at Wuhan University of Technology, Baoding Science and Technology Park, and other locations. Unlike CroHD dataset \citep{9577483}, we recorded and annotated the practical videos by ourselves instead of annotating existing Internet videos. We recorded videos using a 1080P camera in 9 scenes. We recorded videos using a 480P camera only in the classroom scene.  

In this study, due to ethical considerations, we ensured that individuals captured in the video data were fully informed about the purpose, nature, and potential uses of the collected data. (1) In office, roof, and classroom scenarios, we recruited both participants and volunteers and obtained their consent after explaining how their data would be used, how long it would be retained, and the security measures in place to protect their privacy. Participants were paid for their involvement, and consent was obtained prior to any video recording, ensuring that participation was both voluntary and fully informed. (2) In other public area scenarios (such as school roads, streets, school parking lots, etc.), we placed signs in prominent locations to inform the public that video recording was taking place, ensuring transparency and offering opt-out options where applicable. This approach complies with ethical standards for passive data collection in public spaces.

Our Cchead dataset contains 50,528 frames with about 2,366,249 heads and 2,358 tracks. Our Cchead dataset includes 10 different scenes, including indoor and outdoor environments (e.g., streets, classrooms, roofs, school roads, lamps, glass, and parking lots). The movement of crowded pedestrians is diverse, leading to complex congestion and collision avoidance behaviors. The number of people per frame ranges from 11 to 114. Most sequences in Cchead have a framerate of 50 fps. By sampling, high framerates (e.g., 50fps) can decrease to low framerates (e.g., 25fps) to meet practical requirements. We provide high-framerate videos and annotations for flexible and diverse applications. We calculate the mean normalized area of bounding boxes by using $\frac{1}{N} \sum_{i=1}^N\frac{h_{i}*w_{i}}{H*W}$. N is the number of bounding boxes; each box has height and width ($h_{i}$ and $w_{i}$); H and W denote the height and width of the entire image, respectively. We also calculate the moving speed of head by using $\frac{||C_i-C_{i-1}||_2}{Area_{i-1}}$. $C_i$ means the center position of one bounding box at time $i$, and $Area_{i}$ means the bounding box area at time $i$. Since mainstream head detection and tracking methods encompass two perspectives (Slope and Overhead views), Cchead is captured from two viewpoints to facilitate a wider range of application scenarios. The summary of our Cchead dataset is shown in Table~\ref{dataset}. We show some examples in Figure~\ref{datasetexample}. Background objects include vehicles, shops, trees, tables, and chairs, ensuring the diversity of the dataset and facilitating complex deep-learning training.

% Because the moving heads cause significant variations in scale, pose, texture, and illumination, which results in the difficulty in tracking annotation.
\clearpage \newpage 
\begin{sidewaystable}[h!]
\caption{Comparison of Cchead dataset against existing head datasets.}
\centering
\begin{tabular}{llllllll}
% \begin{tabular*}{\textheight}{@{\extracolsep\fill}ccccccccccccccc}
\toprule
Dataset & Head Task & Image & Head box & Scene Number & Scene Description & Source & Resolution \\
\midrule 
HollywoodHeads \citep{ISI:000380414100323} &Detection&224,740&369,846& 21 &Hollywood Movies &Web-scraped&Diverse  \\
SCUT-HEAD \citep{8545068} & Detection & 4,405 & 111,251 & 2 & Classroom & \makecell{Web-scraped\footnotemark[1] (half),\\ In situ\footnotemark[2] (half)} & 1280x720 \\
Brainwash \citep{DBLP:journals/corr/StewartA15} & Detection & 11,917 & 91,146 & 1 & Brainwash Cafe & In situ & 640×480 \\
\textcolor{black}{RPEE-Heads \citep{abubaker2024rpee}} & \textcolor{black}{Detection}& \textcolor{black}{1886} &  \textcolor{black}{109,913}&\textcolor{black}{3}& \textcolor{black}{Railway platforms, entrances, etc.}&\textcolor{black}{Web-scraped}& \textcolor{black}{Diverse}\\
CrowdHuman \citep{shao2018crowdhuman} & Detection & 24,370 & 470K & Unknown & Street, beach, etc. & Web-scraped & Diverse \\
COCOHumanParts \citep{9229236}& Detection & 66,808 & 232,392 & Unknown & Natural scenes  & Web-scraped & Diverse \\

Restaurant \citep{9150687} & Detection & 1,610 & 16,060 & 4 & Restaurant & In situ & 640×480 \\
CroHD \citep{9577483} & \makecell{Detection, \\ \textbf{Tracking}} & 11,463 & 2,276,838 & 5 & Train station, Street & \makecell{Web-scraped (half), \\ In situ (half)}& 1920x1080 \\ \hline
\textbf{Cchead} & \makecell{Detection,  \\ \textbf{Tracking} }& \textbf{50,528} & \textbf{2,366,249} & \textbf{10} & \textbf{School, Road, Roof, etc.} & \textbf{In situ} & \textbf{1920x1080} \\ 
\botrule
\end{tabular}
\footnotetext{Note: }
\footnotetext[1]{They have created an 'In situ' video dataset, captured using cameras on location.}
\footnotetext[2]{They have obtained a 'Web-scraped' video dataset, collected from various online sources.}
\label{compare-dataset}
\end{sidewaystable}

\begin{sidewaystable}[h!]
\caption{Comparison of our methods against other head tracking methods on the Cchead dataset. We train and test these models on our Cchead dataset for fairness.}\label{sotaTRACK}
\tabcolsep=0.5em
\begin{tabular*}{\textheight}{@{\extracolsep\fill}lllllllllllllll}
\toprule%
 \multicolumn{3}{@{}c@{}}{MOT methods}& & & & &\\\cmidrule{1-4}%
        Ref& Detector&ReID&Tracker& IDF1$\uparrow$ & IDs$\downarrow$ & IDP$\uparrow$&IDR$\uparrow$ & MT$\uparrow$ &PT$\downarrow$&ML$\downarrow$ & Rcll$\uparrow$&Prcn$\uparrow$ & MOTA $\uparrow$ \\ 
\midrule
        BoT-SORT \citep{aharon2022bot}&PPYOLOE-L&~&BoT-SORT & 62.4 & 4894 & 66.9&58.5 & 656&352&16 & 80.1&91.8 & 72.0 \\ 
        BoT-SORT \citep{aharon2022bot}&YOLOX&~&BoT-SORT & 69.7 & 2940& 75.6&64.5 & 641&314&69& 78.1&91.6 & 70.4 \\
        DeepSort\citep{wojke2017simple}&PPYOLOE-L&Pplcnet&DeepSort & 69.6 & 2282 & 75.9&64.1 & 580&424&20 & 77.7&92.3 & 70.9 \\ 
        Bytetrack\citep{BS6-WOS:000904116000001}&PPYOLOE-L&~&Bytetracker & 44.6 & 7228 & 46.9&42.5 & 707&303&14 & 81.3&89.9 & 70.9 \\ 
        Bytetrack\citep{BS6-WOS:000904116000001}&YOLOX&Pplcnet&Bytetracker & 39.8 & 12327 & 38.1&41.5 & 762&233&29 & 84.7&77.8 & 58.4 \\ 
        JDE\citep{BS6-INSPEC:20258751}&YOLOv3&Emb Head \footnotemark[1]& SORT& 24.4 & 10459 & 40.0&17.5 & 8&551&157 & 35.1&80.2 & 23.6   \\ 
        Fairmot \citep{BS6-WOS:000692902100001}&CenterNet &Emb Head&SORT& 71.6 & 12107 & 72.3&70.5 & 747&238&38 & 85.1&87.7 & 71.0  \\
        CenterTrack \citep{zhou2020tracking} &CenterNet &~&CenterTracker&  55.3 &7693 &57.8&52.9& 700&299&25& 81.5&89.1 &70.1 \\ 
        ByteTrack\citep{BS6-WOS:000904116000001} & RTDETR & ~ & ByteTracker & 65.9 & 2095 & 69.4 & 62.7& 730 & 263 & 31 & 82.2 & 91.1 & 73.8 \\ 
        OC-SORT\citep{li2020observation} & RTDETR & ~ & OCSORT & 68.7 & 2505 & 71.4 & 66.0 & 736 & 258 & 29 & 82.7 & 89.5 & 72.6 \\
        BoT-SORT \citep{aharon2022bot}&  RTDETR & ~ & BoT-SORT & 70.3 & 1813 & 75.9 & 65.4 & 635 & 347 & 41 & 79.2 & 92.1 & 72.1 \\ \hline
        \textbf{MDFN(Ours)} &CenterNet&Emb Head&SORT&\textbf{77.5}& 6236& \textbf{78.2}&\textbf{76.6}&\textbf{804}&\textbf{184}&36 &\textbf{87.8}&89.7 &\textbf{76.7} \\
\botrule
\end{tabular*}
\footnotetext{Note: }
\footnotetext[1]{Emb Head: Embedding Head for ReID in the JDE framework.}
\end{sidewaystable}

\clearpage

We compare our Cchead dataset with previous datasets in Table~\ref{compare-dataset}. In comparison to existing head detection datasets such as HollywoodHeads, SCUT-HEAD, Brainwash, CrowdHuman, and Restaurant, our Cchead dataset offers both head bounding boxes and tracks. Additionally, while these datasets primarily focus on head detection, our Cchead dataset places special emphasis on head tracking. It should be noted that there is a significant lack of data sets and methods specifically dedicated to head tracking. In contrast to CroHD, which consists of a combination of half 'Web-scraped' and half 'In situ' videos, our Cchead dataset is meticulously recorded and annotated by our own team in practical scenes. Furthermore, our dataset encompasses a wider range of scenes, including schools, roads, classrooms, etc. Moreover, our dataset captures complex pedestrian flows and scenes, such as pedestrian interactions, occlusions, and object interference.
\begin{figure*}[t!]{}
  \centering
 \begin{minipage}{0.49\linewidth}
  \centerline{\includegraphics[width=\textwidth]{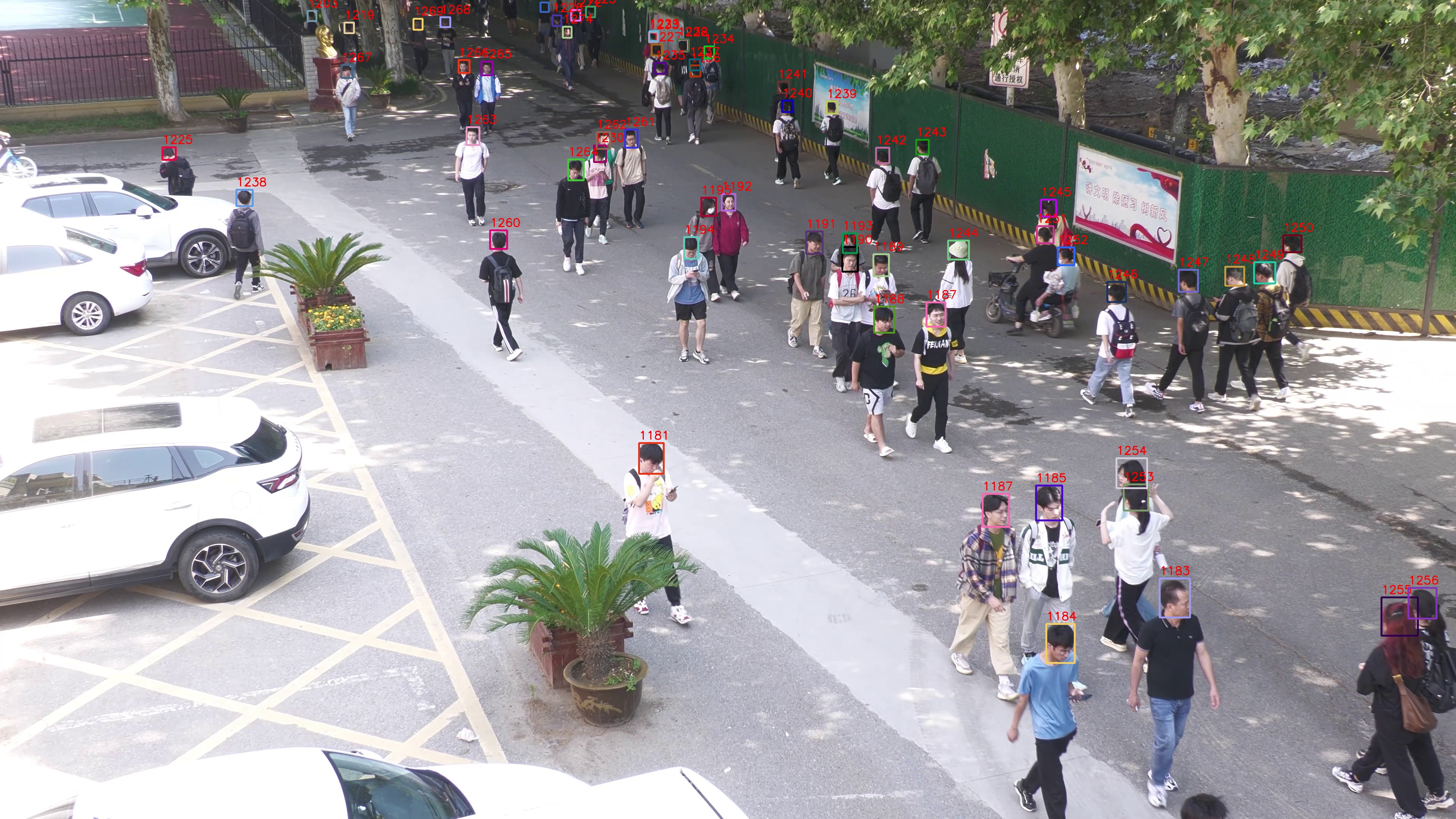}}
  % \centerline{Image}
 \end{minipage}
 \begin{minipage}{0.49\linewidth}
  \centerline{\includegraphics[width=\textwidth]{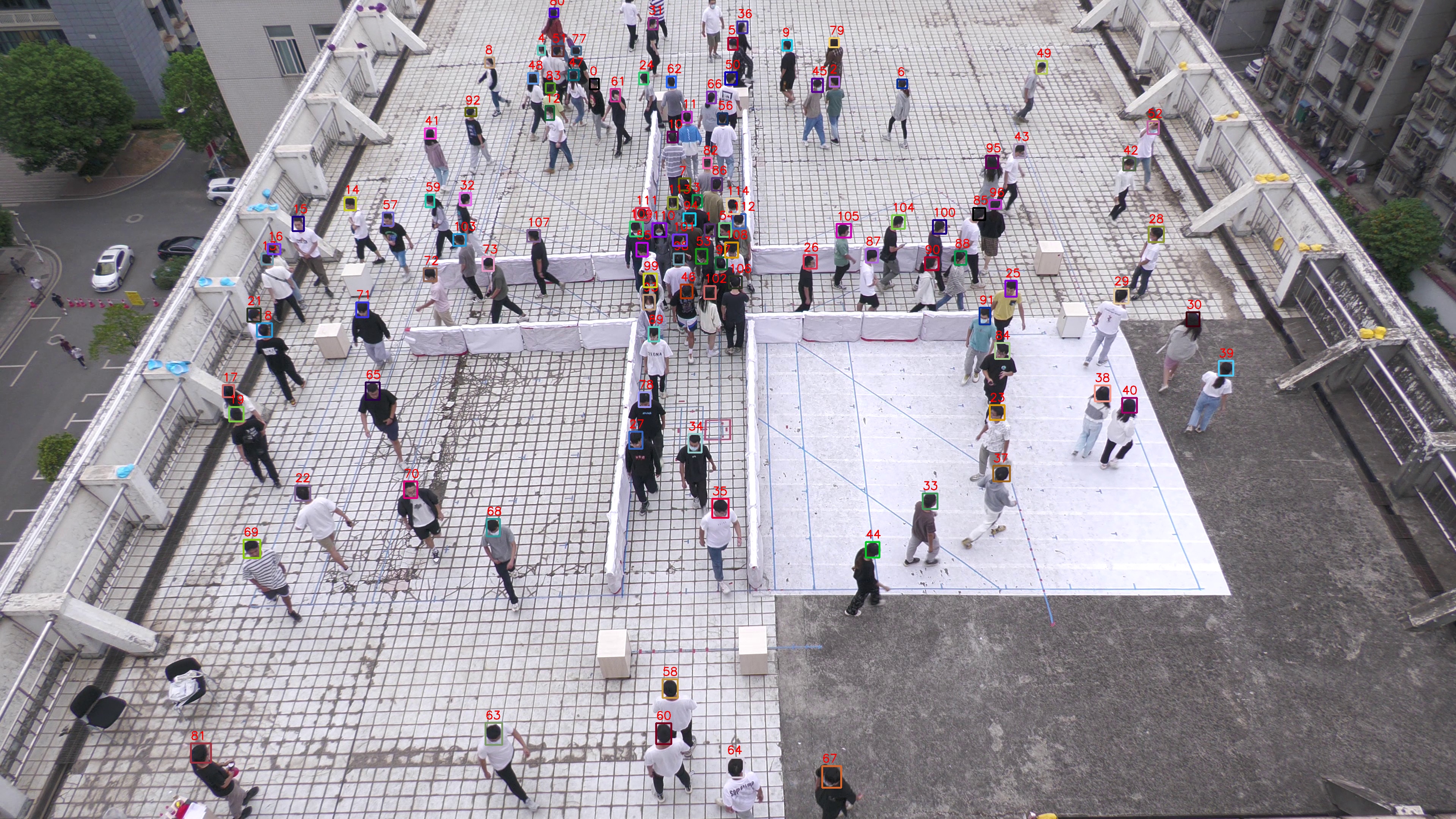}}
 \end{minipage}
  \caption{Examples of the manual annotations for video frames in
our Cchead dataset.}
  \label{example-lsht}
\end{figure*}

\begin{figure}[h]{}
  \centering
  \includegraphics[width=3in]{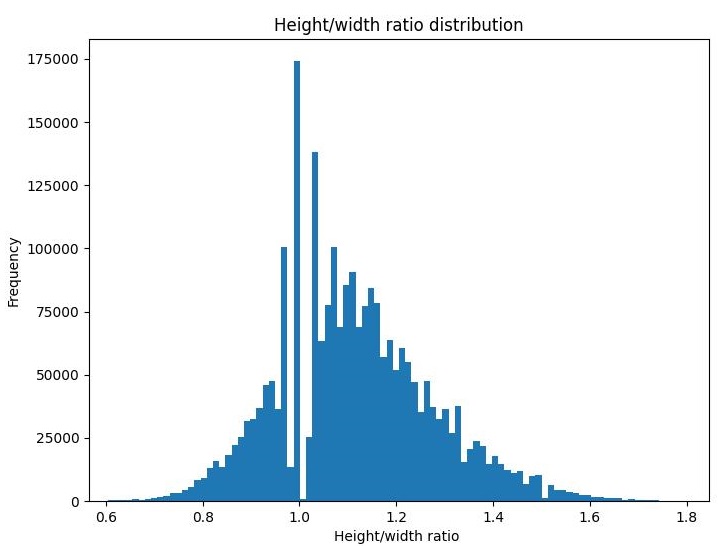}
  \caption{Height/width ratio distribution in annotated head boxes.}
  \label{h1-a}
\end{figure}

We adhere to the format from the MOTChallenge benchmark \citep{milan2016mot16}. The format includes the following fields: box ID, frame, box left, box top, width, height, confidence score, box category, and class visibility. Box ID: A unique identifier for each tracked object. Frame: The frame number in which the object appears. Box left and box top: The top-left x-coordinate and y-coordinate of the bounding box. Width and Height: The width and height of the bounding box. Confidence score: A value representing the confidence level of the detection. Box category: The category or type of the detected object. Class visibility: The visibility ratio of the object within the frame (between 0 and 1). An example:
\begin{quote}
\textit{1, 1, 57, 86, 28, 32, 1, 1, 1 \\
2, 1, 55, 87, 28, 32, 1, 1, 1\\
3, 1, 60, 85, 28, 32, 1, 1, 1\\}
\end{quote}

Our Cchead dataset was annotated using manual annotation and tools such as Darklabel \footnote{https://github.com/darkpgmr/DarkLabel} and computer vision annotation tool (CVAT) \footnote{https://github.com/opencv/cvat}. To ensure the accuracy of the dataset, we adapted manual annotation as the primary method. We employed many annotators and put a lot of effort into annotating accurate pedestrian heads. We meet many challenges in dataset annotation. As for the annotation in an image, the main challenges lie in the small size but diverse poses of the object (head). We have tried many AI auto-annotation tools (e.g., segment anything model \citep{kirillov2023segment}), but the performance is low. 

To develop our high-quality dataset, we initially employed many annotators from a professional annotation company and a university. They spent approximately one month labelling the data to create head boxes and tracks. Following the initial annotation phase, we engaged three experts—comprising two researchers with significant experience in head annotation and one professor in this field—to review the annotated dataset. This process involved two iterative rounds where the dataset annotations were carefully checked and the incorrect annotations were returned to the original annotators for corrections. The cycle (annotation, expert review, and correction) ensured that our dataset achieved a high level of accuracy and reliability. 

To ensure precise annotations for the global research community, we annotated all visible pedestrian heads within each scene, with the visibility score determined based on the annotators' best judgment. Specifically, we consider a proposal to be a match if its Intersection over Union (IoU) with the ground truth exceeds 0.9. This high threshold can ensure that only highly accurate detections are counted as correct matches. As for the annotation in a video, the main challenges lie in that moving heads cause significant variations in occlusion, scale, pose, and texture, resulting in difficulty in annotating trajectory chains. Figure~\ref{example-lsht} exemplifies the resulting annotations of two scenes.

We maintain the same classification of training and test sets for the sequences in Cchead as established by the MOTChallenge benchmark. The dataset is divided into training and testing sets with an approximate 1:1 frame number for each scene, ensuring a balanced distribution for fair evaluation across different scenes. We compute the height/width ratio for each head box in Figure~\ref{h1-a}. It shows that more than half of the head bounding boxes have a height greater than the width. The highest frequency of height/width ratio is 1. The height/width ratio belonging to [0.8, 1.4] is about 93.5\%. We will make our Cchead dataset and annotations publicly available.

The dataset is divided into training and testing sets, with roughly equal frame ratios for each scene,

\textbf{Cchead Audio Dataset.} With the rapid development of deep multimodal learning \citep{RN34553,8103116} and audio-visual correspondence \citep{afouras2022self} in computer vision tasks, audio has become an important input in scene understanding. The audiovisual modalities are complementary to each other. We also collect the audio dataset. It contains 9 scenes: Classroom, Roof(+), Roof(Y), Roof(T), Street, School Road 1, School Road 2, School Parking Lot 1, and School Parking Lot 2. The audio signals, synchronized with the image sequences of our Cchead dataset, are recorded by our cameras. The recorded time of audio signals is the same to our Cchead dataset. To open up future opportunities, we also make our Cchead Audio dataset publicly available.

% For each video, a 1:1 frame ratio was used for partitioning, where the half of the frames were used for training and th remaining half for testing. 

% \begin{table}[h]
% \caption{Caption text}\label{tab1}%
% \begin{tabular}{@{}llll@{}}
% \toprule
% Column 1 & Column 2  & Column 3 & Column 4\\
% \midrule
% row 1    & data 1   & data 2  & data 3  \\
% row 2    & data 4   & data 5\footnotemark[1]  & data 6  \\
% row 3    & data 7   & data 8  & data 9\footnotemark[2]  \\
% \botrule
% \end{tabular}
% \footnotetext{Source: This is an example of table footnote. This is an example of table footnote.}
% \footnotetext[1]{Example for a first table footnote. This is an example of table footnote.}
% \end{table}

% \begin{figure*}[htb]{}
%   \centering
%   \includegraphics[width=5.95in]{SAM.jpg}
%   \caption{Meta-SAM}
%   \label{SAM1}
% \end{figure*}

\section{MDFN}

\begin{figure*}[ht]{}
  \centering
  \includegraphics[width=6.2in]{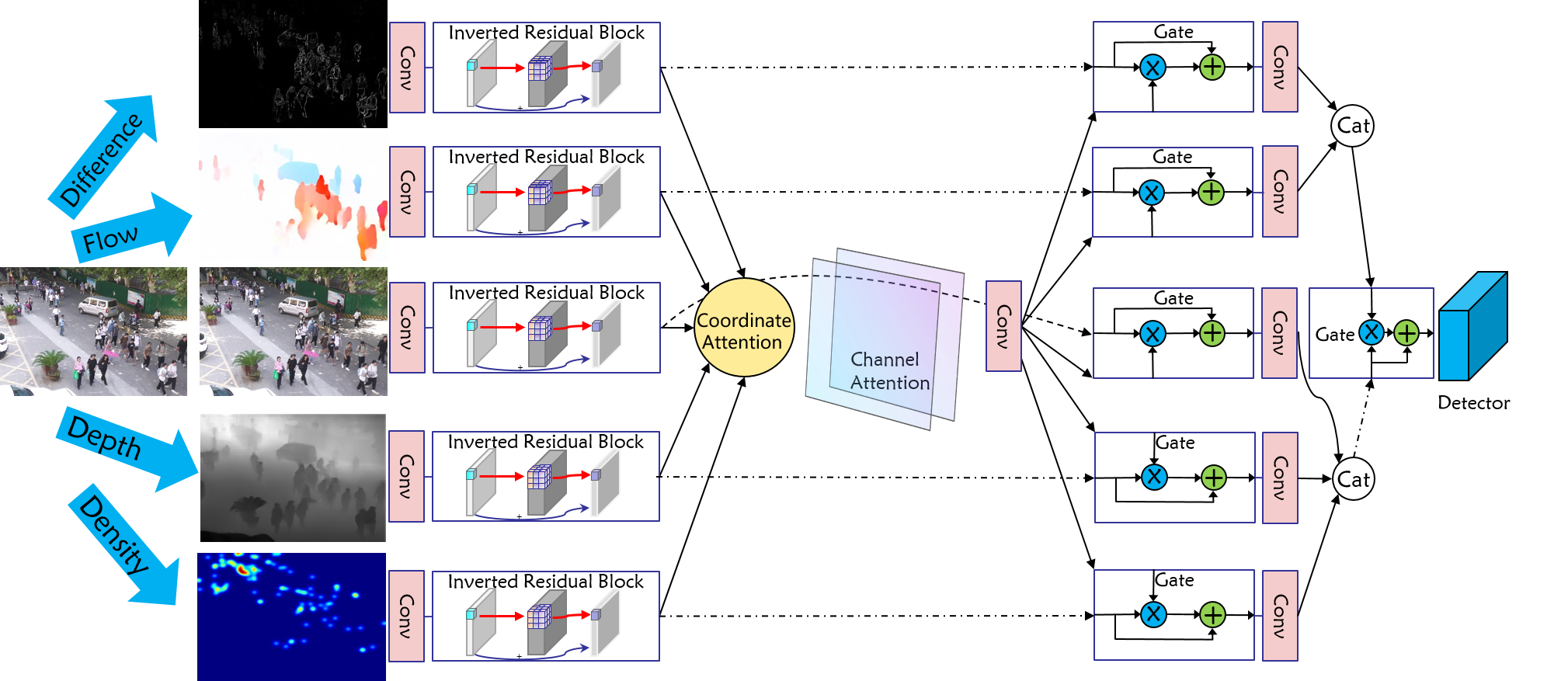}
  \caption{The network structure of MDFN.}
  \label{h1-i}
\end{figure*}

\footnote{Part of MDFN has been used in the 14th International Conference on Applied Human Factors and Ergonomics (AHFE 2023)}The key innovation of this section is applying multi-source data to suppress the background object and enhance the head feature, in the pixel-level head feature space. In general, we have a backbone network $\emph{N}_{{\rm feat}}$ that aims to extract the deep feature, and a detection network $\emph{N}_{{\rm det}}$ that aims to detect the head. The output of the detection network for the input image $\bm{I}$ is $\emph{N}_{{\rm det}}(\bm{h})$, where $\bm{h}$ denotes the extracted feature: $\bm{h}=\emph{N}_{{\rm feat}}( \bm{I})$. 
The inference pipeline is shown in Figure~\ref{h1-i}. It has two
parts: (1) The multi-source data generation module is developed to estimate the five-source information, including optical flow, frame difference maps, depth maps, and density maps.
(2) The feature aggregation module aims to fuse multi-source features for effective head detection and tracking.  

\subsection{Multi-source data generation}
Our MDFN only utilizes the original image $\bm{I}$  without external sensors. Considering the existing computer vision tasks in head areas, we want to make full use of the prior knowledge of heads. Thus, we adopt a transfer learning strategy, using deep neural networks to automatically generate multi-source data and feed them into our feature fusion network in parallel. We will next introduce how to use pre-trained models to obtain them. Following our previous work \citep{SUN2022109354}, we consider both the motion information and static information to be critical to head detection and tracking. We define the motion estimation operator as 

\begin{equation}\begin{aligned}
&\bm{I}_{{\rm diff}}=|\bm{I_f}-\bm{I_{f-1}}|,\bm{f}=1,\dots ,N,\\
&\bm{I}_{{\rm flow}}=\mathcal{F}_{{\rm flow}}(\bm{I_f},\bm{I_{f-1}}),\bm{f}=1,\dots ,N, \\
\end{aligned}
\end{equation}
where $\mathcal{F}_{{\rm flow}}$ is the function of optical flow estimation between the previous frame $\bm{I_{f-1}}$ and the current frame $\bm{I_f}$ in an image sequence. $\mathcal{F}_{{\rm flow}}$ is obtained by an approximation function \citep{BS2-teed2020raft}. $\bm{I}_{{\rm diff}}$ is the function of frame difference between $\bm{I_f}$ and $\bm{I_{f-1}}$. It is directly computed as the difference between these two consecutive frames. The motion information $\mathcal{F}_{{\rm flow}}$ and $\bm{I}_{{\rm diff}}$ could efficiently suppress the static background information while enhancing the head features.

Next, we define the static estimation operator as 
\begin{equation}\begin{aligned}
&\bm{I}_{{\rm dept}}=\mathcal{F}_{{\rm dept}}(\bm{I_f}),\\
&\bm{I}_{{\rm dens}}=\mathcal{F}_{{\rm dens}}(\bm{I_f}),\\
&\bm{I}=\bm{\bm{I_f}}.\\
\end{aligned}
\end{equation}
where $\mathcal{F}_{{\rm dept}}$ is the function of depth estimation. It is used to estimate the depth value of each pixel relative to the camera. The distance patterns of heads are significant, which can help to recognize heads. $\mathcal{F}_{{\rm dept}}$ is obtained by an approximation function \citep{BS6-yuan2022new}. $\mathcal{F}_{{\rm dens}}$ is the function of head density estimation, for highlighting the spatial location of heads. It is obtained by an approximation function \citep{BS6-10.1007}. The static information, $\mathcal{F}_{{\rm dept}}$, $\bm{I_f}$, and $\mathcal{F}_{{\rm dens}}$ are combined for head detection and tracking.

In summary, we generate five-source information, and the aggregated feature can be written as: $\bm{h}=\emph{N}_{{\rm feat}}( \bm{I},\bm{I}_{{\rm diff}},\bm{I}_{{\rm flow}},\bm{I}_{{\rm dept}},\bm{I}_{{\rm dens}})$.

\subsection{Feature aggregation}
We have generated the five-source information, and this section will introduce how to aggregate them. For the convenience of feature fusion, we use CNN to map the five-source information into the feature space. In particular, a pseudo-siamese network—designed with the same architecture but using different weights—is employed to extract features from multiple inputs. This allows the network to learn complementary representations from diverse sources while maintaining structural consistency. After extracting these features, we perform two levels of feature aggregation to effectively combine the information across different feature hierarchies.

At the \textbf{first} aggregation level, we concatenate the five distinct features at the channel dimension. This can ensure that all feature maps are aligned in the same spatial dimensions while expanding the channel dimension:

\begin{equation}
\begin{aligned}
\bm{h}_{{\rm cat}}=&{\rm Cat}(\emph{N}_{{\rm feat1}}(\bm{I}_{{\rm diff}}),\emph{N}_{{\rm feat2}}(\bm{I}_{{\rm flow}}),\\ & \emph{N}_{{\rm feat3}}(\bm{I}),\emph{N}_{{\rm feat4}}(\bm{I}_{{\rm dept}}),\emph{N}_{{\rm feat5}}(\bm{I}_{{\rm dens}})),
\end{aligned}
\end{equation}
where ${\rm Cat}(\cdot)$ is the concatenation function. $\emph{N}_{{\rm featn}}(\cdot)$ is the pseudo-siamese network.

\begin{figure}[h]{}
  \centering
  \includegraphics[width=4in]{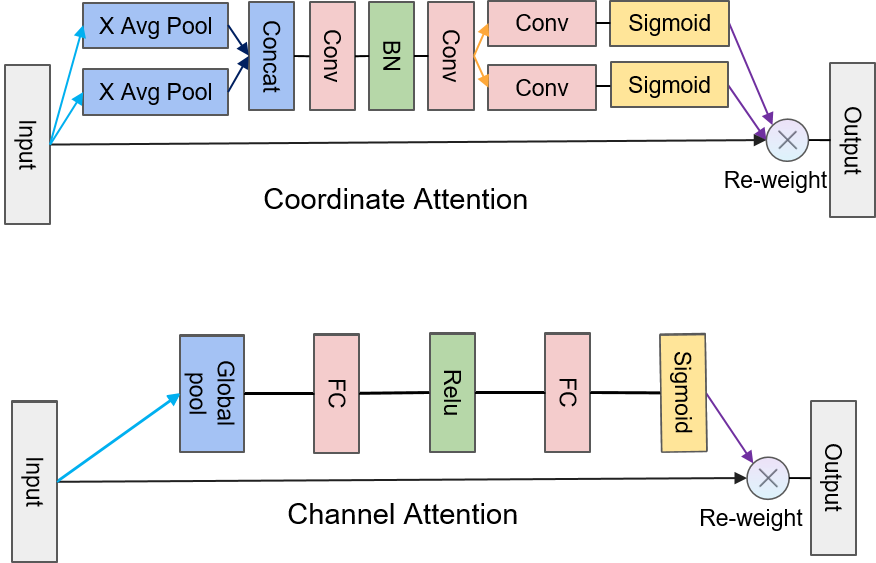}
  \caption{ Coordinate Attention and Channel Attention network \citep{DBLP:journals/corr/abs-2103-02907}.}
  \label{h-fusion}
\end{figure}

Next, we introduce a hybrid strategy that combines convolutional attention with a self-attention mechanism, leveraging the strengths of both to enhance feature representation. For the convolutional attention mechanism, we employ a convolutional network to efficiently capture local dependencies within the feature maps:

\begin{equation}\bm{h}_{{\rm agg}}={\rm ChA}({\rm CoA}({\rm Conv}({\rm Conv}(\bm{h}_{{\rm cat}})))),
\end{equation}
where ${\rm Conv}$ is the standard convolution. ${\rm CoA}$ is the Coordinate Attention network \citep{DBLP:journals/corr/abs-2103-02907}, while ${\rm ChA}$ is the Channel Attention network. We first utilize ${\rm CoA}$ to extract the spatial coordinate information of multi-source coupling. (In Fig.\ref{h-fusion}, \textcolor{black}{${\rm CoA}$ extracts spatial coordinate information by applying average pooling. Feature maps are averaged in horizontal and vertical directions, respectively. These pooled feature maps are then concatenated and passed through a batch normalization (BN) layer. Then  a convolution layer is cascaded to generate re-weighting factors and extract weighted spatial information.}) We next utilize ${\rm ChA}$ to balance channel feature maps with different weights.  \textcolor{black}{(In Fig.\ref{h-fusion}, ${\rm ChA}$ enhances channel-wise feature representation by first applying a global pooling operation to each channel, followed by two fully connected (FC) layers with a ReLU activation. The final sigmoid activation generates channel-wise weights, which are multiplied by the original feature maps to emphasize important channels.}) We convert each feature map into a single value, only keeping the channel information. The single value is considered a channel weight in Fig.\ref{h-fusion}. The channel weights are multiplied by the original feature maps, thus effectively extracting the weighted features.

Next, we will introduce the proposed self-attention method to refine the $\bm{h}_{{\rm agg}}$ by using the original features $\bm{h}_{{\rm cat}}$:
\begin{equation}\label{Hadamard}
\begin{aligned}
\bm{h}_{{\rm agg}}=\alpha_1 \cdot {\rm Sigmod}({\rm Conv}({\rm Conv}(\bm{h}_{{\rm agg}}))) \odot \bm{h}_{{\rm cat}}\\ + \beta_1 \cdot\bm{h}_{{\rm cat}},
\end{aligned}
\end{equation}
where we use the Sigmod function after the aggregated feature ($\bm{h}_{{\rm agg}}$) to construct a learnable special mask (the value of this mask belongs to [0, 1]). This mask focuses on learning the spatial score of the feature maps. This score can represent where features are effective or ineffective. The mask is multiplied with the features by element-wise (Hadamard) product, as defined in Eq.~\ref{Hadamard}. The mask can activate the effective head features while suppressing and filtering the ineffective head features. 

At the \textbf{second} aggregation level, we split the channels into separate groups and then re-concatenate the features:

\begin{equation}
\begin{aligned}
\bm{h}_{{\rm motion}}=&{\rm Cat}({\rm Conv}({\rm Conv}(\bm{h}_{{\rm agg1}})),\\ &{\rm Conv}({\rm Conv}(\bm{h}_{{\rm agg2}})), \\
\bm{h}_{{\rm static}}=&{\rm Cat}({\rm Conv}({\rm Conv}(\bm{h}_{{\rm agg3}})),\\ &{\rm Conv}({\rm Conv}(\bm{h}_{{\rm agg4}})),{\rm Conv}({\rm Conv}(\bm{h}_{{\rm agg5}})),\\
\end{aligned}
\end{equation}
where we apply convolution layers within each group to further refine the extracted information, after separating the channels. Following this, we concatenate the corresponding features related to motion and static information separately. This structured fusion ensures that both dynamic and static aspects are adequately captured, setting the stage for the next aggregation step. 

\begin{equation}
\bm{h}_{{\rm agg}}=\alpha_2 \cdot \bm{h}_{{\rm static}} \odot \bm{h}_{{\rm motion}}+\beta_2 \cdot \bm{h}_{{\rm static}}.
\end{equation}

We multiply the motion information and static information by the element-wise (Hadamard) product. Following our previous work \citep{SUN2022109354}, this operator can effectively improve performance, because the element-wise (Hadamard) product can help to enhance head motion features and suppress background features.

Last, the aggregated features will be applied to detectors and trackers.

\section{Algorithmic Analysis}

Considering head detection is a fundamental task, we train recent SOTA head detection algorithms on our Cchead dataset. Then, MDFN is compared with them in Section~\ref{detect}. Considering head tracking is an advanced task, we compare MDFN with the SOTA MOT algorithms in Section~\ref{track}.

We provide 50fps videos and annotations for flexible and diverse applications. Considering that 25fps is a general framerate parameter in ordinary cameras, all experiments in this study are conducted on the Cchead dataset with 25fps by video sampling. Researchers can adopt other framerate parameters for their practical requirements.

All experiments in this study are conducted on a computer with a Linux system, and 8 NVIDIA 3090 graphics processing units (GPUs). All experiments are based on CUDA version 11.7. The programming language is Python 3, and the deep learning libraries are Paddle 2.4.2, Pytorch 1.9, and PaddleDetection 2.6. 

To evaluate MOT (Multi-Object Tracking) performance, the following six metrics, as detailed in \cite{weblink7} and \cite{ristani2016performance}, are used in the experiments:
\begin{itemize}
    \item MOTA (↑): Multi-Object Tracking Accuracy. 
\item IDP (↑):  The fraction of computed detections that are correctly identified.
\item IDR (↑):  The fraction of ground truth detections that are correctly identified.
\item IDF1 (↑): The ratio of correctly identified detections over the average number of ground-truth and computed detections.
\item MT (↑): Total number of mostly tracked targets, with more than 80\% of their ground-truth trajectories being tracked correctly.
\item ML (↓): Total number of mostly lost targets, with less than 20\% of their ground-truth trajectories being tracked correctly.
\item PT (↓): Total number of targets with 20\%-80\% of their ground-truth trajectories being tracked correctly.
\item Rcll (↑): Ratio of correct detections to the total number of ground truth boxes. 
\item Prcn (↑): Ratio of correct detections to the total number of detection boxes. 
\item IDs (↓): Number of identity switches. 
\end{itemize}
In this context, the notation ↑ (↓) denotes that higher (lower) measurement values signify better tracking performance. Detailed descriptions of these evaluation measures can be found on the official MOT website \citep{weblink7} and \citep{ristani2016performance}.

\subsection{Head detection on Cchead dataset}\label{detect}

To evaluate the SOTA head detection methods, we train and test them on our Cchead dataset in Table~\ref{Ccheaddete}. For evaluation metrics, the standard average precision ($ AP_{50}$) is computed in this study. Faster Region-Based Convolutional Neural Network (Faster R-CNN) \citep{BS4-fastercnn} is a typical two-step object detector with a region proposal network. Then, we use the recent Swin \citep{DBLP:journals/corr/abs-2103-14030} backbone to replace the traditional backbones to improve performance. As one of the important one-stage object detectors, the Center-based Object Detection Network (CenterNet) \citep{9010985} detects the object as a triplet instead of a bounding box. We follow the CenterNet method and use the deep layer aggregation (DLA)-34 \citep{8578353} backbone for head detection. PaddlePaddle (PP)-YOLOE \citep{ppdet2019} is an improved You Only Look Once Version 3 (YOLOV3) variant, which combines various existing tricks to achieve high accuracy while ensuring efficient speed. Vision Transformer (ViT) \citep{10.1007/978-3-031-20077-9_17} backbone is integrated into PP-YOLOE. You Only Look Once Version X (YOLOX) \citep{BS4-yolox} is an anchor-free detector. YOLOX integrates the  YOLO series for higher performance. We use  Cross Stage Partial Network (CSPResNet) \citep{9150780} as the backbone of YOLOX. Detection Transformer (DETR) simplifies detection pipelines by employing a transformer for object sequence prediction. To enhance its performance with small object detection, Deformable DETR \citep{zhu2020deformable} introduces a deformable attention module, effectively boosting detection accuracy for smaller objects. Real-time Detection Transformer (RT-DETR) \citep{10657220} improves the inference speed of DETR with an efficient hybrid encoder and uncertainty-minimal query selection strategy.

Compared with SOTA methods, our MDFN achieves the best performance on our Cchead dataset. Our MDFN, using only pseudo multi-source input (optical flow, depth, frame difference, and density maps) without external sensors, is more accurate than YOLOX with a 0.06 $ AP_{50}$ boost.

\begin{figure*}[htbp]
% \begin{center}
 \begin{minipage}{0.49\linewidth}
  \centerline{\includegraphics[width=\textwidth]{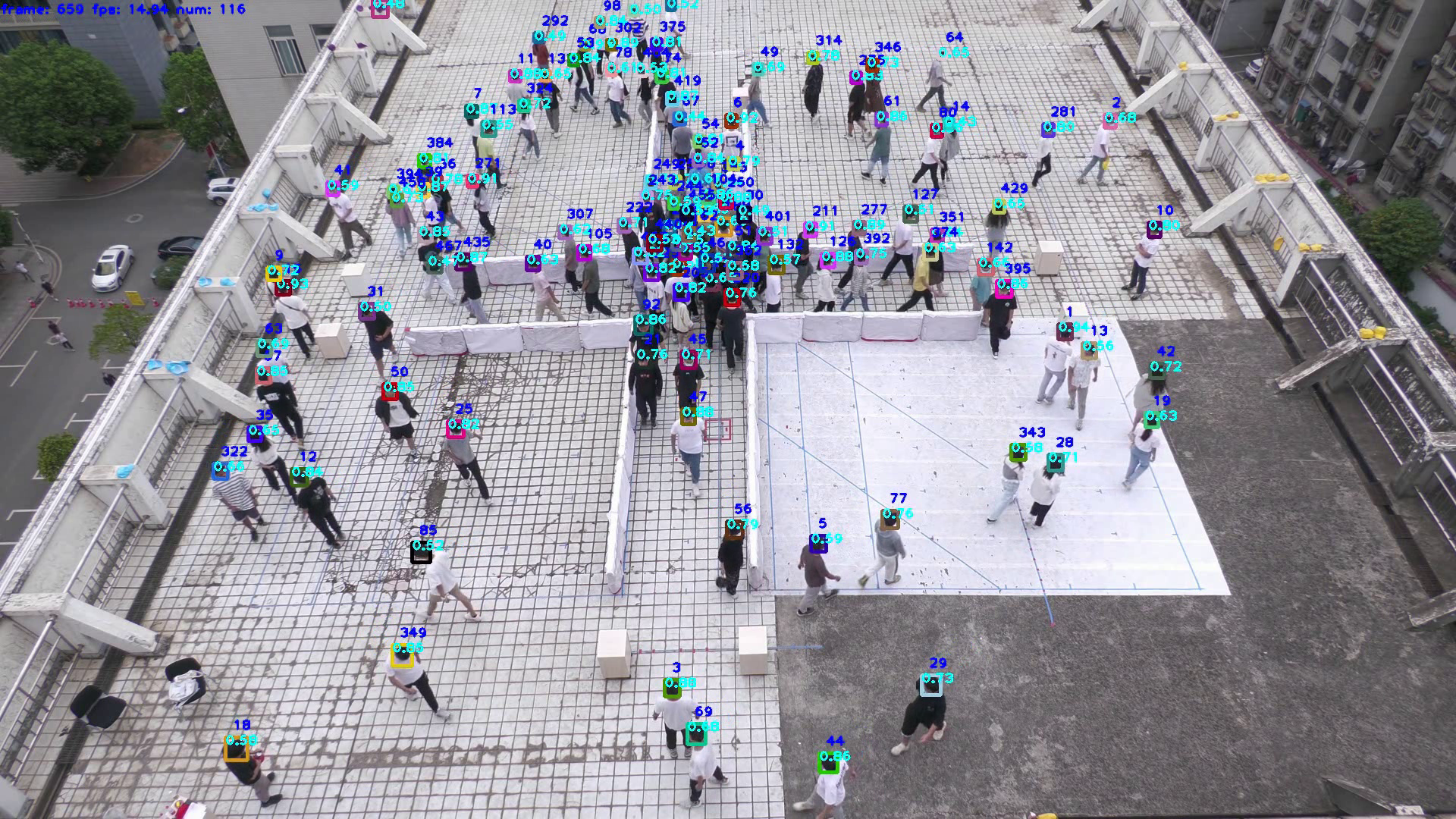}}
  % \centerline{Image}
 \end{minipage}
 \begin{minipage}{0.49\linewidth}
  \centerline{\includegraphics[width=\textwidth]{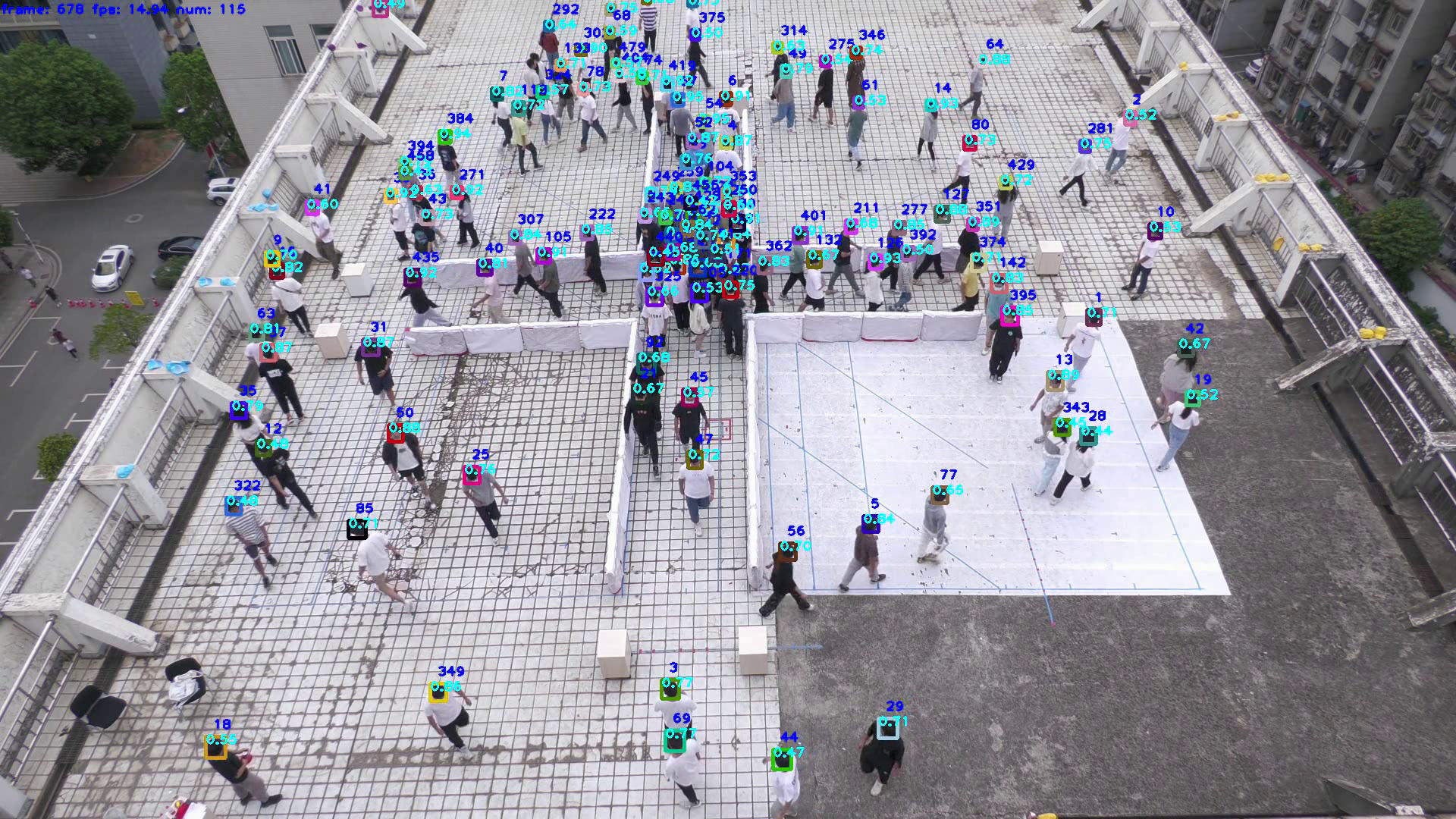}}
 \end{minipage}
 % \begin{minipage}{0.33\linewidth}
 %  \centerline{\includegraphics[width=\textwidth]{c4.jpg}}
 % \end{minipage}

 % \begin{minipage}{0.33\linewidth}
 %  \centerline{\includegraphics[width=\textwidth]{c7.png}}
 % \end{minipage}
 \begin{minipage}{0.49\linewidth}
  \centerline{\includegraphics[width=\textwidth]{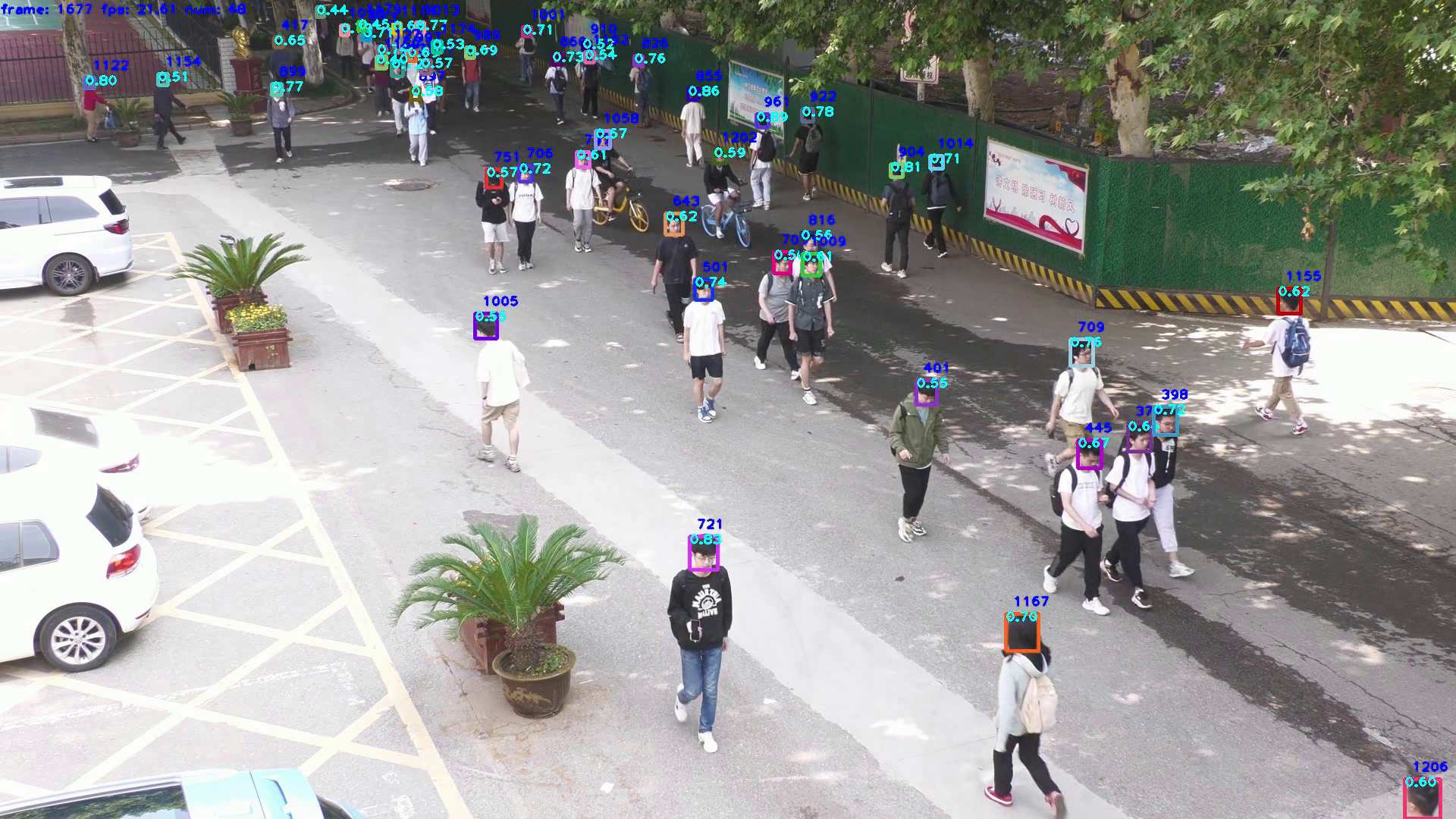}}
 \end{minipage}
 \begin{minipage}{0.49\linewidth}
  \centerline{\includegraphics[width=\textwidth]{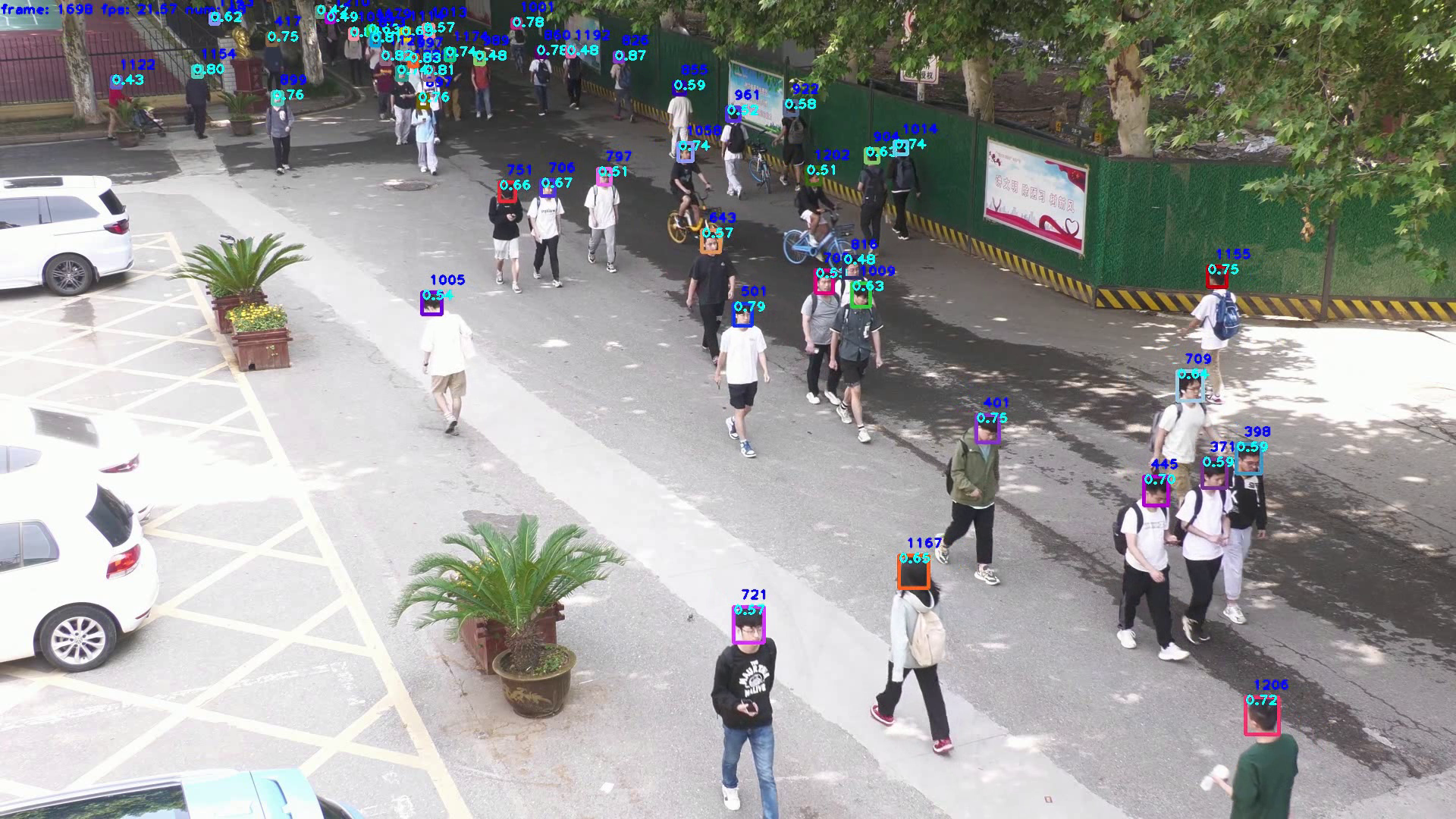}}
 \end{minipage}
 
 \begin{minipage}{0.49\linewidth}
  \centerline{\includegraphics[width=\textwidth]{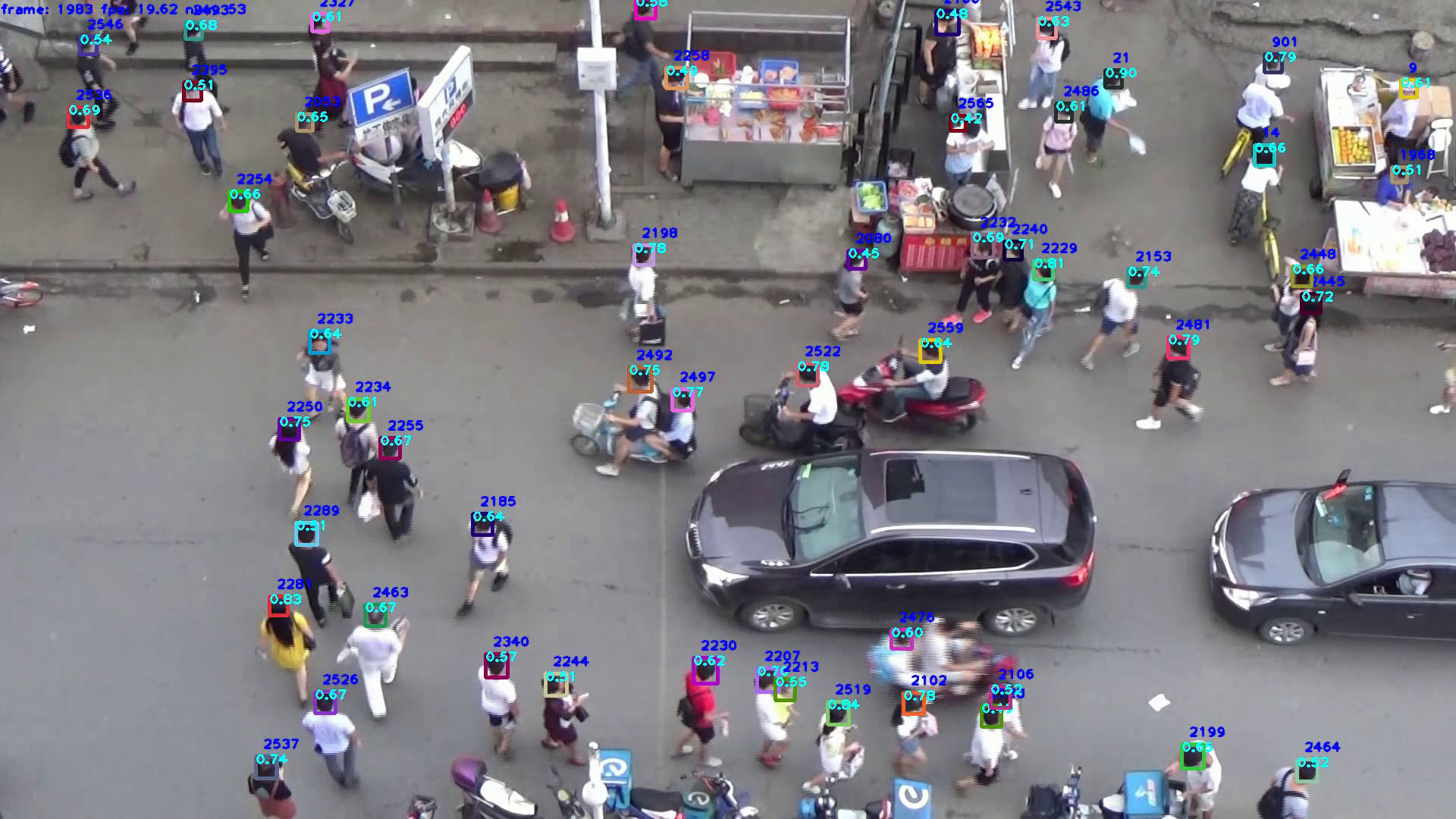}}
 \end{minipage}
 \begin{minipage}{0.49\linewidth}
  \centerline{\includegraphics[width=\textwidth]{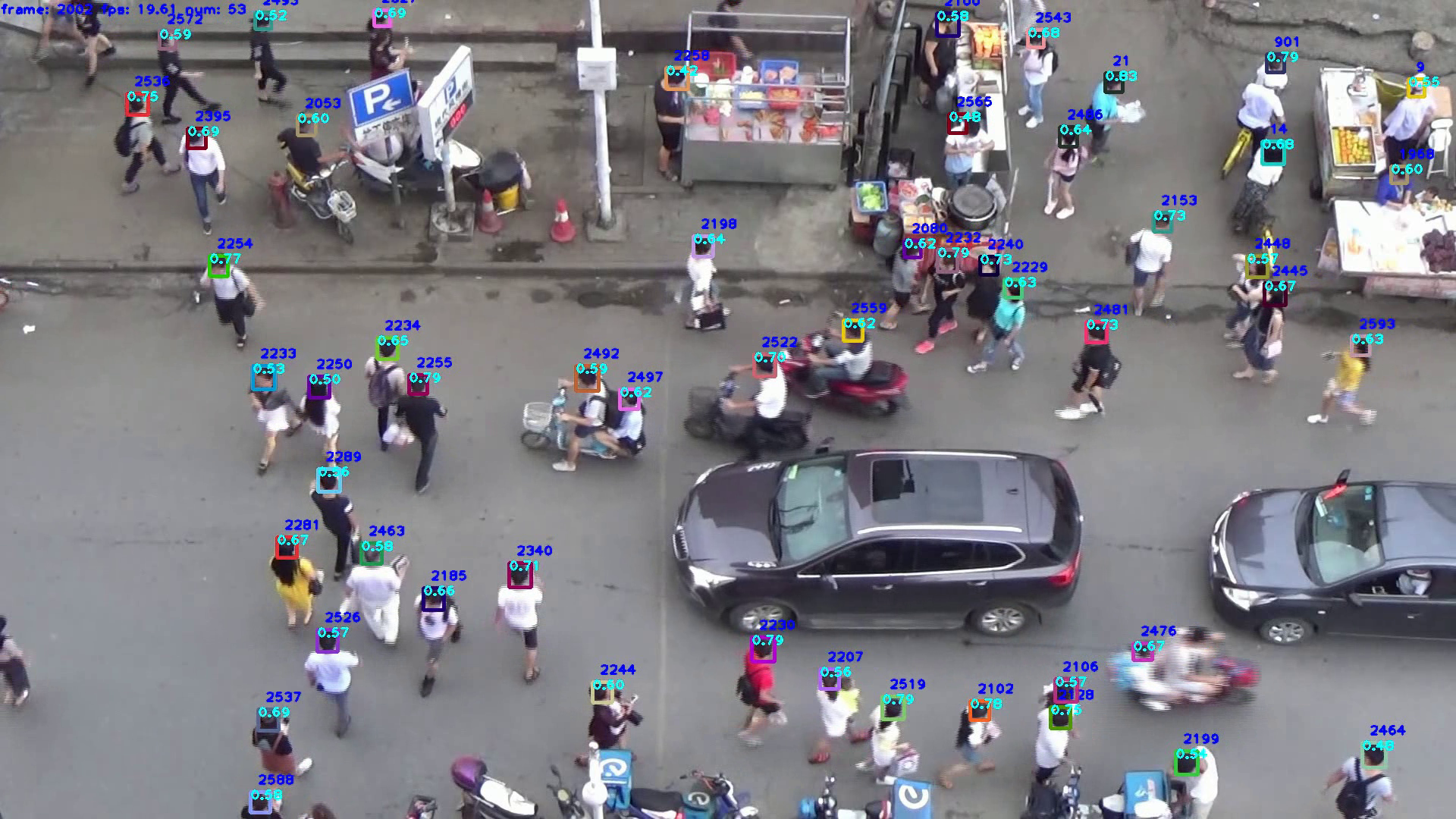}}
 \end{minipage}
 
 \hspace{10 em} Frame 1     \hspace{17 em}   Frame 20
\caption{Qualitative visualization of our MDFN on the Cchead testset with various crowded scenes. Blue numbers indicate IDs of pedestrians. More visualization videos can be found in \protect\footnotemark.}
\vspace{-0.5em}
\label{vis-example}
\end{figure*}
\footnotetext{\url{https://drive.google.com/drive/folders/1BLmzCRx3MbOzVUITw0-RCpRqTHJ2JXYQ?usp=sharing}}

\begin{table}[t]
\caption{Comparison of MDFN against other SOTA head detection methods on Cchead dataset. We train and test these models on the same dataset for fairness.}
\vspace{-1em}
\begin{tabular}{@{}p{8cm}p{3cm}p{3cm}@{}}
\toprule
 Method&Backbone&$ AP_{50}$\\
\midrule
 Faster R-CNN \citep{BS4-fastercnn}&Swin  &0.81 \\
 Centernet \citep{9010985}&DLA-34& 0.67 \\
 PP-YOLOE \citep{ppdet2019} & ViT & 0.76 \\
 YOLOX-X \citep{BS4-yolox} &CSPResNet&0.81 \\
 Deformable DETR \citep{zhu2020deformable} &ResNet50& 0.80 \\
% MPSN (Sun et al., 2022) MobileNetv2 0.84 \\
 RT-DETR \citep{10657220}&ResNet101&0.82 \\\hline
 \bf{MDFN (Ours)} &CSPResNet& \bf{0.86} \\
\bottomrule
\end{tabular}
\vspace{-1em}
\label{Ccheaddete}
\end{table}

\subsection{\textcolor{black}{Head detection on Restaurant dataset}}\label{detect-rest}

\textcolor{black}{
To further evaluate MDFN, we test the publicly available crowd Restaurant dataset \citep{9150687}. The Restaurant dataset was collected in four different indoor locations at a restaurant. It includes 1610 images, from which the test set contains 123 images. The images are extracted from the video at a large time interval, thus creating significant diversity and difference. }

\textcolor{black}{
Our detector is based on FCHD \citep{Vora2018FCHDAF}. The training hyper-parameters are given. Backbone uses the first 11 layers of the MobileNetv2 network pre-trained on the ImageNet dataset. The anchor box sizes are selected as 2 and 4, and the whole model is trained by the SGD optimizer for 50 epochs. The learning rate is $10^{-2}$, which decays to $10^{-3}$, $10^{-4}$, and $10^{-5}$ after 15, 35, and 42 epochs, respectively. The detector RPN network consists of 5 convolutional layers, initialized using a standard normal distribution with a standard deviation of 0.01. We set $\alpha_1=\beta_1=\alpha_2=\beta_2=1$. The batch size is set to 1.}

\begin{figure}[ht]{}
  \centering
  \includegraphics[width=5in]{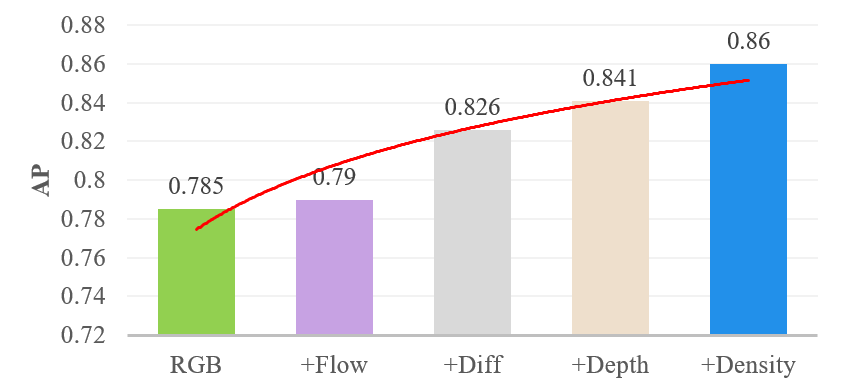}
  \caption{\textcolor{black}{Ablation study on the Restaurant testset. Testing results of MDFN w.r.t. the multi-source information. MN2: MobileNetv2.}}
  \label{abla_MDFN}
\end{figure}

\begin{table}[t]
\caption{\textcolor{black}{Comparison of MDFN against other SOTA methods on Restaurant dataset. We train and test these models on the same dataset for fairness.}}
\vspace{-1em}
% \begin{tabular}{@{}ccc@{}}
\begin{tabular}{@{}p{8cm}p{3cm}p{3cm}@{}}
\toprule
 Method&Backbone&$ AP_{50}$\\
\midrule
 HTC++ \citep{DBLP:journals/corr/abs-1901-07518} &Swin-B& 0.68 \\
 CrowdDet \citep{Chu_2020_CVPR}& ResNet50& 0.61 \\
 SSD \citep{Zhang_2018_CVPR} &ResNet18 &0.51 \\
 FCHD \citep{Vora2018FCHDAF} &VGGNet16 &0.75 \\
 Iter-E2EDET \citep{2022arXiv220307669Z} &ResNet50& 0.77 \\ \hline
% MPSN (Sun et al., 2022) MobileNetv2 0.84 \\
 \bf{MDFN (Ours)} &MobileNetv2& \bf{0.86} \\
\bottomrule
\end{tabular}
\vspace{-1.7em}
\label{rest}
\end{table}

% YOLOX \citep{DBLP:journals/corr/abs-2107-08430} is a high-performance anchor-free detector through integrating YOLO series. 

% \footnote{https://paperswithcode.com/sota/object-detection-on-crowdhuman-full-body} so far. 

\textcolor{black}{
The results are shown in Table~\ref{rest}. SSD \citep{Zhang_2018_CVPR} algorithm uses multi-scale feature maps to achieve high detection accuracy HTC++ framework \citep{DBLP:journals/corr/abs-1901-07518} combines detection and segmentation tasks into a joint multi-stage processing method and utilizes spatial context to further boost the performance.
CrowdDet \citep{Chu_2020_CVPR} is a SOTA detector that achieves 0.907 AP performance in the challenging CrowdHuman dataset. 
Iter-E2EDET \citep{2022arXiv220307669Z} is a SOTA people detector and achieves very high accuracy in the CrowdHuman detection task. Our MDFN is superior to other SOTA algorithms (including detectors based on CNN and Transformer). It is worth noting that we use the lightweight MobileNetv2 network. Studies show the lightweight MobileNetv2 network will decrease object detection performance \citep{8578572}. Even in this case, our algorithm is still excellent.
For further analysis, we conduct ablation experiments \citep{AYDIN2022103661} to explore the importance of multi-source information, as shown in Figure~\ref{abla_MDFN}. We find that: the pseudo-multi-source information effectively improves the network performance and does not require external sensors; motion information, especially frame difference information, plays a key role in head detection; density information can further improve and enhance the head detection network. }

\subsection{Head Tracking on Cchead dataset}\label{track}

In this section, we train and compare recent SOTA multiple object tracking models \citep{weblink7} on our Cchead dataset. For instance, Bytetrack \citep{BS6-WOS:000904116000001} is a two-stage tracking algorithm (SDE): it first detects the object boxes in each frame, dividing them into two classes: high-confidence and low-confidence boxes, and retains the high-confidence boxes while designing rules to recover tracking trajectories for the low-confidence boxes. In data association, Bytetrack employs Kalman filtering to estimate the future moving trajectory, and then utilizes IoU to compute similarity matching. Both FairMOT \citep{BS6-WOS:000692902100001} and JDE \citep{BS6-INSPEC:20258751} are one-step tracking algorithm (JDE). FairMOT utilizes an encoder-decoder structure to simultaneously perform object detection and ReID tasks. FairMOT uses the ResNet-34 backbone network and CenterNet as the detector. FairMOT considers ReID feature extraction as a classification task and adds this ReID branch from the CenterNet.

The results compared in the test set are in Table~\ref{sotaTRACK}. We follow the standard evaluation metrics from \citep{web-link8}. To make the comparison fair, we adopt the same strategy for traning Fairmot and MDFN. In particular, We apply Heatmap, Offset and Size loss functions for the detection task and Crossentropy loss function for the ReID task. 

We train the MDFN model using the Momentum optimizer for 30 epochs. The learning rate is $5^{-3}$ and decline to $5^{-4}$ and $5^{-5}$ in 15th and 22nd epoch. The ReID network contains two convolutional layers. 

Apart from JDE and Bytetrack, existing other algorithms perform closely. Our MDFN significantly outperforms SOTA MOT algorithms, achieving a 76.7 MOTA. Our MDFN is based on the SOTA Fairmot algorithm with a 5.9 ID F1 score (IDF1) improvement. We do not need external sensors. Moreover, We improve over the SOTA algorithm by a nontrivial margin in the highly competitive benchmark.

\begin{figure}[htbp]
% \begin{center}
 \begin{minipage}{0.49\linewidth}
  \centerline{\includegraphics[width=\textwidth]{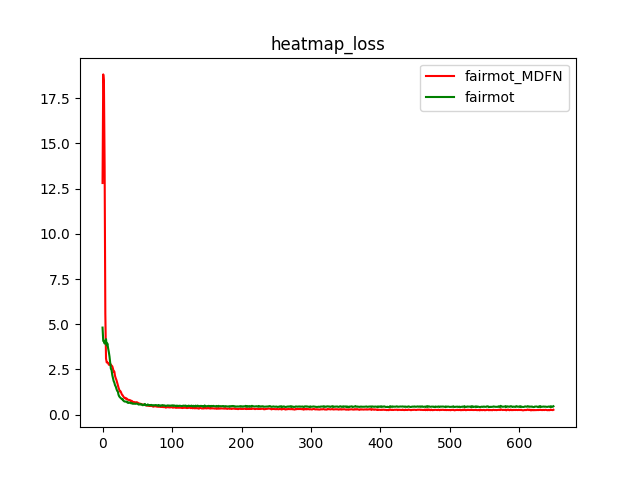}}
  % \centerline{Image}
 \end{minipage}
 \begin{minipage}{0.49\linewidth}
  \centerline{\includegraphics[width=\textwidth]{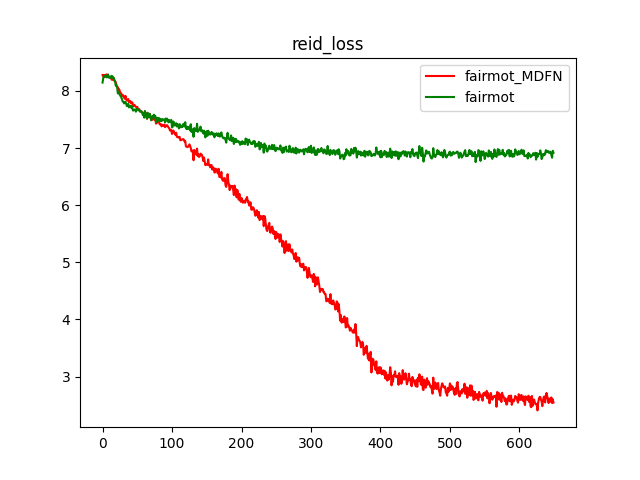}}
 \end{minipage}
 \begin{minipage}{0.49\linewidth}
  \centerline{\includegraphics[width=\textwidth]{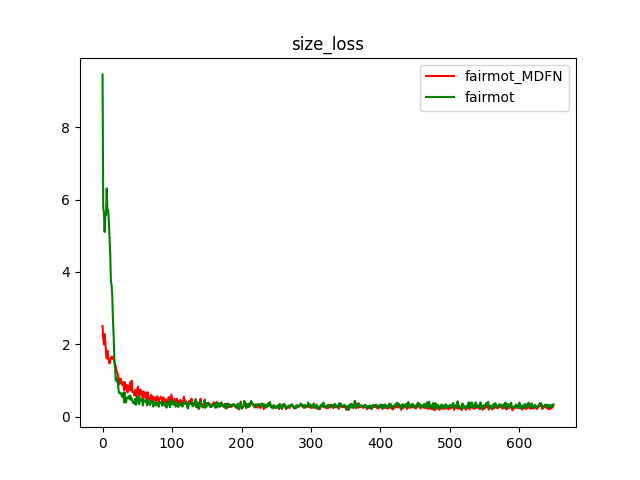}}
 \end{minipage}
 \begin{minipage}{0.49\linewidth}
  \centerline{\includegraphics[width=\textwidth]{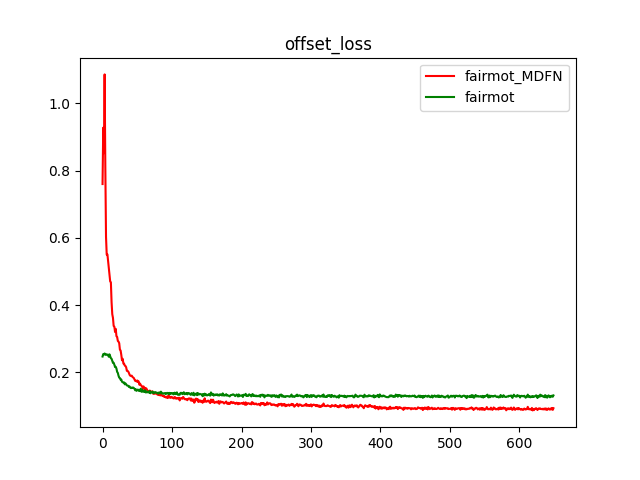}}
 \end{minipage}
\caption{Comparison of MDFN with the baseline method (Fairmot). Training losses of MDFN w.r.t. Training times (./20 batchsizes).}
\vspace{-0.5em}
\label{lossplot}
\end{figure}

\textbf{MDFN v.s. Fairmot.}
In the original Fairmot paper \citep{BS6-WOS:000692902100001},  the training losses include two parts: the detection and the Re-ID branchs. They are different and in charge of different tasks (detection and tracking) .The former (detection branch) includes three parts: heatmap head aims to estimate the locations of the object centres with focal loss; offset and size heads aim to localize objects more precisely in box offset and size with $l_1$ losses; Re-ID head aims to extract features and distinguish objects through a classification task with crossentropy loss. We visualize the training processes of MDFN and Fairmot in Figure~\ref{lossplot}. The corresponding metric evaluation of trained MDFN is in Table~\ref{sotaTRACK}. 

We find that the four losses of our MDFN converge to lower values than Fairmot's. In particular, for the Re-ID loss, our MDFN's loss (red line) achieves a very low value (about 2.59), compared with Fairmot (about 6.90). This comparison shows that our MDFN can help the head tracking task more.

% Full size image
The qualitative visualization of our MDFN is illustrated in Figure~\ref{vis-example}. The performances are evaluated on different crowded scenes (e.g., Roof, School Road). We visualize many positive examples where pedestrian heads are accurately tracked , even though heads are small and crowded. Besides, we provide similar visualization videos of other SOTA methods (e.g., BoT-SORT, DeepSort, Fairmot, and Bytetrack) in \footnote{\url{https://drive.google.com/drive/folders/1m-ZA9rPey-DJpuSrs67tnj-GRqoFisWS?usp=sharing}}.

In summary, these above observations and comparisons show that our MDFN can boost head tracking performance. To our knowledge, multi-source data with feature fusion is not often used in head tracking studies.

\subsection{Head Tracking on CroHD dataset}\label{CroHD-RESULTS}
\textcolor{black}{In this section, we train and compare recent SOTA multiple object tracking models on the CroHD dataset \citep{9577483}.   The CroHD dataset comprises 11,463 images across nine Full-HD resolution sequences, collected through web scraping and in situ captures. Five of these sequences are derived from the publicly available MOTChallenge CVPR19 benchmark \citep{Dendorfer2020MOT20AB}. The dataset covers five distinct scenes, including train stations and streets (daytime and nighttime). We divide the CroHD dataset into training and testing sets with an 8:2 ratio. Tab. \ref{sota-TRACK} presents the comparative results. Our MDFN significantly outperforms state-of-the-art (SOTA) MOT algorithms, achieving an IDF1 of 75.9, an IDP of 78.4, and a MOTA of 64.1. Ablation experiments in Tab. \ref{sota-TRACK} show that the use of multi-source data leads to improvements of 3.9 in IDF1, 4.4 in IDP, 3.4 in IDR and 5.3 in MOTA. Additionally, MDFN method does not rely on external sensors. In summary, substantial improvements are achieved over SOTA algorithms in this competitive benchmark, demonstrating the strong generalization ability of our MDFN algorithm across multiple datasets.}

\begin{table}[h!]
\caption{Comparison of our methods against other head tracking methods on the CroHD dataset. We train and test these models on our CroHD dataset for fairness. }\label{sota-TRACK}
\setlength\tabcolsep{2pt}
\begin{tabular}{l|llllllll|lll}
\toprule%
        \multirow{2}{*}{MOT method} & 
        \multicolumn{8}{l}{Tracking} &\multicolumn{3}{l}{Detection}\\ 
        & IDF1$\uparrow$ & IDs$\downarrow$ & IDP$\uparrow$&IDR$\uparrow$ & MT$\uparrow$ &PT$\downarrow$&ML$\downarrow$ & MOTA $\uparrow$&Rcll$\uparrow$&Prcn$\uparrow$ & F1 $\uparrow$ \\ 
\midrule
        % YOLOX+Bytetrack & 63.4\% & 4178 & 64.1\%&62.7\% &1579 & 993&138 &81.1 \% & 83.0\% &64.1\%  \\ 
        PPYOLOE+DeepSort \citep{wojke2017simple} & 50.7\%& 2918&55.3\%&46.8\%&467&457&79&58.0\%&72.0\%&85.1\%&77.9\%\\
        PPYOLOE+Bytetrack \citep{BS6-WOS:000904116000001} & 49.6\%& 2638&56.0\%&44.5\%&421&473&109&55.5\%&68.1\%&85.7\%&75.8\%\\
        YOLOX+Bytetrack  \citep{BS6-WOS:000904116000001} & 54.2\%& 2099&59.3\%&49.8\%&467&443&93&60.0\%&72.5\%&\textbf{86.3\%}&78.8\%\\
        YOLOX+BotSort  \citep{aharon2022bot}& 65.9\% & 1523 & 74.9\% & 58.8\% & 419 & 483 & 101 & 55.6\% & 67.4\% & 85.8\% & 75.6\% \\  
        PPYOLOE+BotSort \citep{aharon2022bot} & 52.0\% & 2803 & 54.6\% & 49.6\% & 518 & 424 & 68 & 57.1\% & 74.6\% & 82.2\% & 78.2\% \\  
        \hline
        \hline
        % MDFN(Ours)  RGB & 70.4\% & 1568 & 72.5\% & 68.4\% & 463 & 469 & 71 & 75.7\% & 80.3\% & 56.4\% \\  
        \textbf{MDFN}(RGB) & 72.0\% & 1418 & 74.0\% & 70.1\% & 494 & 446 & \textbf{63}&58.8\% & 77.1\% & 81.4\% & 79.2\% \\  
        \textbf{MDFN}(RGB+Diff+Flow) & 73.9\% & 1816 & 75.5\% & 72.4\% & 522 & 414 & 67& 61.1\% & 78.9\% & 82.3\%&80.6 \% \\  
        \textbf{MDFN}(RGB+Diff+Flow+Depth+Density) & \textbf{75.9}\% & \textbf{1289} & \textbf{78.4\%} & \textbf{73.5\%} & \textbf{530} & \textbf{384} & 89 & \textbf{64.1\%}& \textbf{79.3\%} & 84.5\% &\textbf{81.7\%} \\  
\botrule
\end{tabular}
\end{table}

\section{Practical Implications}

\textcolor{black}{
The proposed Cchead dataset and MDFN algorithm provide many practical implications for real-world applications. First, as a diverse and richly annotated benchmark, the Cchead dataset can fill the gap in complex, head-level pedestrian tracking—particularly in crowded and occluded scenes.  Our dataset has features that are of considerable interest, including 10 diverse scenes of 50,528 frames with over 2,366,249 heads and 2,358 tracks annotated. The dataset captures a wide range of pedestrian movement patterns, including varying speeds, directions, and complex collision-avoidance behaviors.  What's more, all data in our Cchead dataset were collected from real-world scenes and meticulously annotated by professional annotators to ensure the high quality. This provides a solid foundation for developing and evaluating advanced computer vision algorithms in crowd scenarios such as subways, streets, parking lots, schools, and public squares. }

\textcolor{black}{
Second, the multi-source data fusion strategy enhances tracking robustness by integrating motion information, spatial density, and depth information without requiring external sensors. MDFN is suitable for robust deployment in surveillance systems and edge computing platforms. We provide a comprehensive analysis and comparison with existing state-of-the-art (SOTA) algorithms, where MDFN achieves superior performance on Restaurant, CroHD and Cchead datasets.}

\textcolor{black}{
Third, our dataset and method can facilitate crowd management and urban planning by offering fine-grained pedestrian analysis. For instance, the annotated trajectories and movement speeds in Cchead can facilitate behavioral modeling essential for emergency evacuation planning, smart city infrastructure design, and public safety monitoring. Moreover, MDFN can autonomously track indoor occupants, highlighting the potential for smart building management systems to improve both safety and energy efficiency. To encourage further research and development, we publicly release the Cchead dataset, source code, and trained models to the global research community. We believe this work will facilitate innovation and applications across interdisciplinary domains, including computer vision, intelligent transportation, urban planning and smart cities.}

\section{Conclusion}\label{sec13}
 
\textcolor{black}{
We presented the Chinese large-scale cross-scene pedestrian head-tracking dataset (Cchead), including 10 diverse scenes of 50,528 frames with over 2,366,249 heads and 2,358 tracks annotated. {The dataset is collected from real-world scenarios, making it highly applicable in practical applications while also providing researchers with broader opportunities for further exploration. On the other hand,}  we proposed MDFN, which aggregates multi-source data from a transfer learning strategy and achieves superior performance {across three datasets: Cchead, Restaurant and CroHD}.In particular, MDFN achieves the highest tracking performance, i.e., 77.5 IDF1 and 76.7 MOTA on Cchead dataset, 75.9 IDF1 and 78.4 IDP on the CroHD dataset. Ablation experiments confirm the significance of pseudo multi-source data fusion (e.g., optical flow, density and depth map) without external sensors. We believe that the release of the Cchead dataset and our open-source MDFN model will serve as a foundation for advancing research in pedestrian analysis, smart city monitoring, and crowd modeling.}

\textcolor{black}{
This study has two limitations. First, although MDFN introduces feature-level fusion, the temporal modeling of long-term tracking could be further improved with the pseudo multi-source data. Second, while the Cchead dataset includes synchronized audio, the MDFN does not fully explore multimodal learning potentials involving audio-visual fusion. }

\textcolor{black}{
In the future, deploying real-time, lightweight pedestrian trakcing models on edge devices is an important and promising direction. This could involve applying model compression techniques such as network pruning, quantization, and knowledge distillation \citep{SUN2025128791}. Network pruning reduces parameters by removing redundant connections, convolutional kernels, or channels. Model quantization decreases model size and accelerates inference by converting 32-bit floating-point weights into lower-precision formats such as 16-bit floats or 8-bit integers. Knowledge distillation transfers the learned representations from a large teacher model to a smaller student model, enabling the latter to approximate the performance of the former. On the other hand, another direction is to expand the head tracking dataset to include a broader range of lighting conditions, environmental contexts, and weather scenarios. This should include low-light environments such as nighttime scenes or poorly lit indoor areas \citep{ghari2024pedestrian}, which pose challenges for standard RGB-based head detection and tracking methods. Moreover, extreme weather conditions—such as heavy rain, dense fog, and snow—can significantly degrade visual quality and tracking reliability. By collecting data under such diverse conditions, these datasets could better support robust pedestrian tracking algorithms and generalize well to real-world deployments.}

% On the other hand, in indoor applications, privacy concerns are particularly critical. To mitigate this issue, future work could consider replacing RGB cameras with privacy-preserving sensors such as infrared (IR) and millimeter-wave radar, which provide more user-friendly and ethically sound solutions for human detection and tracking.

% incorporating audio-visual multimodal fusion to enhance tracking in scenes with high acoustic variability;

% failure modes of these algorithms. In this section we discuss some of the key lessons we learned over the years of
% ILSVRC, strive to address the key criticisms of the datasets
% and the challenges we encountered over the years, and conclude by looking forward into the future.

\backmatter

% \bmhead{Supplementary information}

% If your article has accompanying supplementary file/s please state so here. 

% Authors reporting data from electrophoretic gels and blots should supply the full unprocessed scans for key as part of their Supplementary information. This may be requested by the editorial team/s if it is missing.

% Please refer to Journal-level guidance for any specific requirements.

\bmhead{Data Availability Statement} The proposed dataset is available for reviewers and editors first. After the peer review, the proposed dataset will be available to global researchers.
% wishing to use it for non-commercial purposes.

\bmhead{Acknowledgments}

% This work is supported by National Natural Science Foundation of China under Grant No. 62192751 and the 111 International Collaboration Program of China under Grant No. BP2018006.

This work was sponsored by grants from National Natural Science Foundation of China (Grant No. 52172308, 62192751), Knowledge Innovation Program of Wuhan (Grant NO. 2023010201010093) and Department of Transport
of Hubei Province (Contract No. 2024-81-3-10).
\bibliography{sn-article}% common bib file
% \bibliographystyle{unsrt}
% \bibliography{}
\clearpage

\end{document}